\documentclass[journal]{IEEEtran}

\usepackage{amsfonts,amssymb}
\usepackage{graphicx}
\usepackage{cite}
\usepackage{amssymb,amsfonts}
\usepackage{setspace}
\usepackage{textcomp}
\usepackage{color,soul}
\usepackage{multirow}
\usepackage{cuted}
\usepackage{array}
\usepackage{epsfig}
\usepackage{graphics}
\usepackage{varwidth}
\usepackage{amsmath}
\usepackage{breqn}
\usepackage{caption}
\usepackage{subcaption}
\usepackage{breqn}
\usepackage[T1]{fontenc}
\usepackage{mathtools}
\usepackage{amsthm}
\usepackage{mathrsfs}
\usepackage{algorithm2e}
\usepackage{cleveref}
\allowdisplaybreaks
\usepackage{amsmath}

\DeclareMathOperator*{\argmin}{arg\,min}

\newtheorem{remark}{Remark}

\newtheorem{pro1}{Proposition}

\begin{document}

\title{STAR-RIS-Aided Mobile Edge Computing: Computation Rate Maximization with Binary Amplitude Coefficients}

\author{Zhenrong Liu,~\IEEEmembership{Graduate Student Member,~IEEE,}
        Zongze Li,
        Miaowen Wen,~\IEEEmembership{Senior Member,~IEEE,}\\
         Yi Gong,~\IEEEmembership{Senior Member,~IEEE,}
        and Yik-Chung~Wu,~\IEEEmembership{Senior Member,~IEEE}

\thanks{Zhenrong Liu is with the Department of Electrical and Electronic Engineering,
The University of Hong Kong, Hong Kong, and also with the Department of Electrical and Electronic Engineering, Southern University of Science and Technology, Shenzhen 518055, China. Zongze Li is with Peng Cheng Laboratory, Shenzhen 518038, China. Miaowen Wen is with the School of Electronic and Information Engineering, South China University of Technology, Guangzhou 510640, China. Yi Gong is with the Department of Electrical and Electronic Engineering, Southern University of Science and Technology, Shenzhen 518055, China. Yik-Chung Wu is with the Department of Electrical and Electronic Engineering, The University of Hong Kong, Hong Kong.}
}

\markboth{TCOM,~Vol.~X, No.~X, X~2023}%
{Liu \MakeLowercase{\textit{et al.}}: STAR-RIS-Aided Mobile Edge Computing: Computation Rate Maximization with Binary Amplitude Coefficients}

\IEEEpubid{0000--0000/00\$00.00~\copyright~2023 IEEE}

\onecolumn

{\fontsize{14}{16}\selectfont IEEE Copyright Notice}

\

\copyright~2023 IEEE. Personal use of this material is permitted. Permission from IEEE must be obtained for all
other uses, in any current or future media, including reprinting/republishing this material for advertising
or promotional purposes, creating new collective works for resale or redistribution to servers or lists, or
reuse of any copyrighted component of this work in other works.

\newpage
\twocolumn
\maketitle
\begin{abstract}
In this paper, simultaneously transmitting and reflecting (STAR) reconfigurable intelligent surface (RIS) is investigated in the multi-user mobile edge computing (MEC) system to improve the computation rate. Compared with traditional RIS-aided MEC, STAR-RIS extends the service coverage from half-space to full-space and provides new flexibility for improving the computation rate for end users. However, the STAR-RIS-aided MEC system design is a challenging problem due to the non-smooth and non-convex binary amplitude coefficients with coupled phase shifters. To fill this gap, this paper formulates a computation rate maximization problem via the joint design of the STAR-RIS phase shifts, reflection and transmission amplitude coefficients, the receive beamforming vectors, and energy partition strategies for local computing and offloading. To tackle the discontinuity caused by binary variables, we propose an efficient smoothing-based method to decrease convergence error, in contrast to the conventional penalty-based method, which brings many undesired stationary points and local optima. Furthermore, a fast iterative algorithm is proposed to obtain a stationary point for the joint optimization problem, with each subproblem solved by a low-complexity algorithm, making the proposed design scalable to a massive number of users and STAR-RIS elements. Simulation results validate the strength of the proposed smoothing-based method and show that the proposed fast iterative algorithm achieves a higher computation rate than the conventional method while saving the computation time by at least an order of magnitude. Moreover, the resultant STAR-RIS-aided MEC system significantly improves the computation rate compared to other baseline schemes with conventional reflect-only/transmit-only RIS.
\end{abstract}


\section{Introduction}

\IEEEPARstart{I}{n} traditional cloud computing systems, the mobile users' data would be sent to a remote cloud center distant from the end devices\cite{armbrust2010view}. However, this paradigm cannot fit the future Internet of Things era due to the unprecedentedly increasing amount of data generated by a massive number of users and the latency-critical requirements of new applications, such as industrial monitoring multi-sensory communications and mobile virtual reality\cite{7123563}. Therefore, mobile edge computing (MEC) has emerged and has drawn much attention from both academia and industry. It promotes using computing capabilities at the edge servers attached to the wireless access points (APs), where the users' data can be offloaded to the nearby APs, and the results are delivered back to the users after the computation\cite{8016573}. It dramatically reduces latency and communication overhead of the network backhaul, thus overcoming the critical challenges for materializing beyond 5G. 

In the MEC paradigm, the system objective shifts from purely maximizing the communication sum rate to maximizing the computation rate. Since the edge server has powerful computation capability and the users' computation results are in very small sizes, the bottleneck of the computation rate for wireless-aided MEC systems is the uplink offloading performance\cite{8264794,7762913,8334188}. However, the uplink offloading performance is severely restricted in practice due to the energy-limited mobile users and the adverse wireless channel condition. 

\IEEEpubidadjcol

In order to improve the uplink offloading performance in wireless-aided MEC systems, great attention has been drawn to the emerging technology of reconfigurable intelligent surfaces (RIS) due to its advantages of low cost, easy deployment, and directional signal enhancement or interference nulling\cite{8910627,9360709,9722713,9770199,9457078}. However, the traditional RIS can only reflect or transmit the incident wireless signal  (conventional reflect-only/transmit-only RISs). In this case, the AP and users must be located on the same side (for reflect-only RIS) or the opposite side (for transmit-only RIS) of the RIS, leading to only a half-space coverage. This geographical restriction may not always be satisfied in practice and severely limits the flexibility of the deployment and effectiveness of the RISs. To overcome this limitation, the simultaneously transmitting and reflecting RIS (STAR-RIS) was proposed\cite{9690478,9437234,9491943}, where the wireless signals to the STAR-RIS from either side of the surface are divided into two parts \cite{zhu2014dynamic}. One part (reflected signal) is in the same half-space (i.e., the reflection space), and the other (transmitted signal) is transmitted to the opposite half-space (i.e., the transmission space). By controlling a STAR-RIS element's electric and magnetic currents, the transmitted and reflected signals can be reconfigured via two generally independent coefficients, known as the transmission and the reflection coefficients \cite{9437234}. With the recently developed prototypes resembling STAR-RISs\cite{zhu2014dynamic,123,9722826,9895224}, a highly flexible full-space coverage will become a reality. Due to the full-space coverage advantage, the STAR-RIS-aided MEC network will break the geographical restriction of the accessed devices to provide a high coverage ratio and support continuous MEC service with better quality.

In this paper, we aim to study the computation rate of a STAR-RIS-aided MEC system. In particular, we formulate the computation rate maximization problem by considering the local and offloaded computation rates under the energy budget. Since the STAR-RIS's transmission and reflection coefficient matrix couples with the design of receive beamforming vectors and uplink transmission power, the resulting optimization problem is an intertwined design of resource allocation in the STAR-RIS-aided MEC network. Worse still, the reflection and transmission amplitude coefficients of STAR-RIS are discrete binary variables, leading to a challenging mixed-integer non-convex optimization problem. 

To address the above challenges, we transform the binary constraints into their equivalent continuous form and then resort to the penalty-based method. However, this equivalent transformation may introduce many undesired stationary points and local optima\footnote{These undesired stationary points and local optima have a lower objective value than the global optimum.}, thus diminishing the solution quality. Therefore, we further adopt the logarithmic smoothing function for binary variables in the penalized objective to eliminate the undesirable stationary points and local optima. With the above ideas, the resultant problem can be solved under the block coordinate descent (BCD) framework. Furthermore, to obtain an overall low complexity algorithm, each subproblem is solved with either closed-forms or first-order algorithms, with complexity orders only scaling linearly with respect to the number of elements in the STAR-RIS or users. Extensive simulation results show that the proposed STAR-RIS-aided MEC system outperforms systems with conventional RIS. At the same time, the proposed smoothing-based method for binary variables optimization leads to better performance than the conventional penalty-based method.

The rest of the paper is organized as follows. The system model and the computation rate maximization problem are formulated in Section \ref{sym}. The binary reflection and transmission amplitude coefficients design problem are handled in Section \ref{3a}. Section \ref{3} optimizes the other variables under the BCD framework. Simulation results are presented in Section V. Finally, a conclusion is drawn in Section VI.

\textit{Notations}: We use boldface lowercase and uppercase letters to represent vectors and matrices, respectively. The transpose, conjugate, conjugate transpose, and diagonal matrix are denoted as $(\cdot)^{\mathrm{T}},(\cdot)^{*},(\cdot)^{\mathrm{H}},$ and $\operatorname{diag}(\cdot)$, respectively. The symbols $|\cdot|$ and $\operatorname{Re}(\cdot)$ denote the modulus and the real component of a complex number, respectively. The $n \times n$ identity matrix is denoted as $\boldsymbol{I}_{n}$ and the $(i,j)^{th}$ element of a matrix  $\boldsymbol{X}$ is denoted by $(\boldsymbol{X})_{i j}$. The complex normal distribution is denoted as $\mathcal{C N}$. Notations $\leq$ or $\geq$ are used for element-wise comparison. For example, if $\boldsymbol{x}_1 \leq \boldsymbol{x}_2$, $\boldsymbol{x}_2$ is element-wise greater than $\boldsymbol{x}_1$.

\section{System Model and Problem Formulation}
\label{sym}

\subsection{System Model}
\begin{figure}[t]
\centering
\centerline{\includegraphics[scale=0.10]{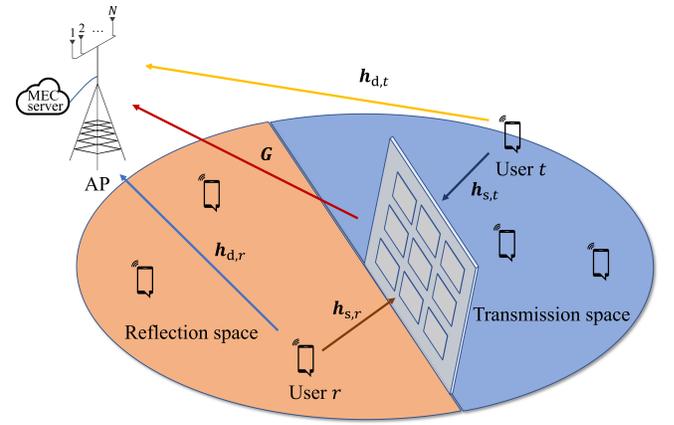}}
\caption{The STAR-RIS-aided MEC system.}
\label{sm}
\end{figure}

We consider a STAR-RIS-aided MEC system as shown in Fig. \ref{sm}, in which there are an AP with $N$ antennas, $K$ single-antenna users, and an $M$-element STAR-RIS. The AP is attached to a MEC edge server. Since the STAR-RIS can provide full-space coverage by allowing simultaneous transmission and reflection of the incident signal, it serves both $T$ users in the transmission space and $R$ users in the reflection space.

Each user has a limited energy budget but has intensive computation tasks to deal with, and the available energy for computing the task in user $k$ is denoted as $E_{k}$ in Joules (J). In practice, if the computation task is too complicated to be completed by a user (possibly due to excessive energy or time involved), part of the task should be delegated to the edge server. Hence, a partial offloading mode is adopted \cite{8016573}. Partial offloading is designed to cope with computation tasks that can be arbitrarily divided to facilitate parallel operations at users for local computing and the AP for edge computing. Being grid-powered, the MEC server has strong computation and storage capabilities for helping users to compute their offloaded tasks and report the computation results in a negligible time.\footnote{In practice, as long as the overall computation workload of users doesn't exceed the MEC server's computing capacity, the computation time can be disregarded as the computation at MEC servers is conducted simultaneously with the uplink data transmission\cite{7762913,8894168,9241750}.}

For uplink transmission, let $\mathcal{T}=\{1, \ldots, T\}$, $\mathcal{R}=$ $\{T+1, \ldots, K\}$ and $\mathcal{K}=\{\mathcal{T}\cup \mathcal{R}\}$ denote the index sets of the $T$ users in transmission space, the $R=K-T$ users in reflection space, and all the users, respectively. Let $s_{t}$ and $s_{r} \in \mathbb{C}$ with zero mean and unit variance denote the information symbols of user $t \in \mathcal{T}$ and user $r \in \mathcal{R}$ for the offloading task, and $p_{t}$ and $p_{r}$ denote the transmit power of user $t$ and user $r$, respectively. Note that all the users with offloading requirements transmit their signals simultaneously, and thus we can express the corresponding received signal $\boldsymbol{y} \in \mathbb{C}^{N\times 1}$ at the AP as
\begin{equation}
\label{first}
\boldsymbol{y}=\sum_{t \in \mathcal{T}} \boldsymbol{g}_{t} \sqrt{p_{t}} s_{t}+\sum_{r \in \mathcal{R}} \boldsymbol{g}_{r} \sqrt{p_{r}} s_{r}+\boldsymbol{z},
\end{equation}
where $\boldsymbol{g}_{t}$, $\boldsymbol{g}_{r}$ $\in \mathbb{C}^{N \times 1}$ are respectively the equivalent baseband channels from the user $t$ and $r$ to the AP, and $\boldsymbol{z} \sim \mathcal{C N}\left(\mathbf{0}, \sigma^{2} \boldsymbol{I}_{N}\right)$ is
the receiver noise at the AP with $\sigma^{2}$ being the noise power. 

With the deployment of a STAR-RIS, the equivalent baseband channels from the user $t$ or $r$ to the AP consists of both the direct link and transmitted or reflected link from the STAR-RIS. Therefore, $\boldsymbol{g}_{t}$ can be modeled as
\begin{equation}
\label{T-channel}
\boldsymbol{g}_{t}=\boldsymbol{h}_{\mathrm{d}, t}+\left(\boldsymbol{G}\right)^{\mathrm{H}}\boldsymbol{\Omega}_t\boldsymbol{\Theta} \boldsymbol{h}_{\mathrm{s}, t},  \quad \forall t \in \mathcal{T},
\end{equation}where $\boldsymbol{h}_{\mathrm{d},t} \in \mathbb{C}^{N \times 1}, \boldsymbol{h}_{\mathrm{s}, t} \in \mathbb{C}^{M \times 1}$, and $\boldsymbol{G} \in \mathbb{C}^{M \times N}$ are the narrow-band quasi-static fading channels from user $t$ to the AP, from user $t$ to the STAR-RIS, and from the STAR-RIS to the AP, respectively. The STAR-RIS phase shift matrix $\boldsymbol{\Theta}=\operatorname{diag}\left(e^{j\theta_{1}}, \ldots, e^{j\theta_{M}}\right) \in \mathbb{C}^{M \times M}$ is a diagonal matrix with $\theta_{m} \in[0,2 \pi)$ being the phase shift of the $m^{th}$ element and $m \in \mathcal{M}=\{1,2, \ldots, M\}$. The STAR-RIS transmission amplitude coefficient matrix $\boldsymbol{\Omega}_t=\operatorname{diag}\left(\rho^{\mathrm{t}}_1, \ldots, \rho^{\mathrm{t}}_M\right) \in \mathbb{R}^{M \times M}$ is a diagonal matrix with $\rho^{\mathrm{t}}_m$ being the transmission amplitude coefficient of the $m^{th}$ element.

Similar to (\ref{T-channel}), $\boldsymbol{g}_{r}$ is expressed as
\begin{equation}
\label{R-channel}
\boldsymbol{g}_{r}=\boldsymbol{h}_{\mathrm{d}, r}+\left(\boldsymbol{G}\right)^{\mathrm{H}}\boldsymbol{\Omega}_r\boldsymbol{\Theta}\boldsymbol{h}_{\mathrm{s}, r},  \quad \forall r \in \mathcal{R},
\end{equation}where $\boldsymbol{h}_{\mathrm{d},r} \in \mathbb{C}^{N \times 1}$, $\boldsymbol{h}_{\mathrm{s}, r} \in \mathbb{C}^{M \times 1}$ denote the narrow-band quasi-static fading channels from the user $r$ to the AP, and from the user $r$ to the STAR-RIS, respectively. The STAR-RIS reflection amplitude coefficient matrix $\boldsymbol{\Omega}_r=\operatorname{diag}\left(\rho^{\mathrm{r}}_1, \ldots, \rho^{\mathrm{r}}_M\right) \in \mathbb{R}^{M \times M}$ is a diagonal matrix with $\rho^{\mathrm{r}}_m$ being the reflection amplitude coefficient of the $m^{th}$ element. In this paper, we consider the mode switching (MS) protocol, where $\rho_{m}^{\mathrm{t}}, \rho_{m}^{\mathrm{r}} \in\{0,1\}$ and $\left(\rho_{m}^{\mathrm{t}}\right)^2+\left(\rho_{m}^{\mathrm{r}}\right)^2=1$. Such an “on-off” type operating protocol is much easier to implement compared to the energy splitting (ES) protocol and therefore more practical in the application scenarios\cite{9690478}. Moreover, from the users' perspective, the MS protocol has lower latency than time switching (TS) protocols, which separates two orthogonal time slots for reflection and transmission users and inevitably incurs higher latency for users assigned to the second time slot.

For the offloading task, we introduce an energy partition parameter $a_{k} \in[0,1]$ for user $k \in \mathcal{K}$, and $a_{k} E_{k}$ represents the energy used for computation offloading. Correspondingly, the transmit power of user $k$ for computation offloading is given as
\begin{equation}
    p_{k}=\frac{a_{k} E_{k}}{L}, \quad \forall k \in \mathcal{K},
\end{equation}
where $L$ is the length of the time slot.

We consider the linear beamforming strategy and denote $\boldsymbol{v}_{k} \in$ $\mathbb{C}^{N \times 1}$ as the receive beamforming vector of the AP for decoding $s_{k}$. Based on (\ref{first}), the received signal at the AP for user $k$, denoted by ${\hat{s}}_{k} \in \mathbb{C}$, is then given by
\begin{equation}
{\!\hat{s}}_{k}\!\!=\!\!\left(\boldsymbol{v}_{k}\right)\!^{\mathrm{H}}\!\boldsymbol{g}_{k} \sqrt{p_{k}}s_{k}\!+\!\left(\boldsymbol{v}_{k}\right)\!^{\mathrm{H}}\!\sum_{l \neq k} \!\boldsymbol{g}_{l}\sqrt{p_{l}}s_l\!+\!\left(\boldsymbol{v}_{k}\right)^{\mathrm{H}} \!\boldsymbol{z}, \ \  \forall k \! \in \! \mathcal{K}.
\end{equation}

The uplink signal-to-interference-plus-noise ratio (SINR) observed at the AP for user $k$ is thus given by
\begin{equation}
\label{upsinr}
\!\!\gamma_{k}\Big(\!\boldsymbol{a}, \boldsymbol{v}_{k},\boldsymbol{\Theta},\boldsymbol{\rho}\!\Big)\!\!=\!\!\frac{p_{k}\!\left|\left(\boldsymbol{v}_{k}\right)^{\mathrm{H}}\!\boldsymbol{g}_{k}
\right|^{2}}{\sum_{l \neq k} \!p_{l}\!\left|\left(\boldsymbol{v}_{k}\right)^{\mathrm{H}}\!\boldsymbol{g}_{l}
\right|^{2}\!\!\!+\!\sigma^{2}\!\left\|\boldsymbol{v}_{k}\right\|^{2}}, \ \  \forall k \!\in\! \mathcal{K},
\end{equation}
where we denote an energy partition vector $\boldsymbol{a}=\left[a_1, \ldots, a_K\right]^{\mathrm{T}}$ and a reflection and transmission amplitude coefficient vector $\boldsymbol{\rho}=[\rho_{1}^{\mathrm{t}},\cdots,\rho_{M}^{\mathrm{t}},\rho_{1}^{\mathrm{r}},\cdots,\rho_{M}^{\mathrm{r}}]^{\mathrm{T}}$.
Then the computation rate of user $k$ due to offloading is
\begin{equation}
\label{rkrho}
R_{k}\!\Big(\!\boldsymbol{a}, \boldsymbol{v}_{k},\boldsymbol{\Theta},\boldsymbol{\rho}\!\Big)\!\!=\!\!B\log _{2}\!\Big(\!1+\gamma_{k}\Big(\boldsymbol{a}, \boldsymbol{v}_{k},\boldsymbol{\Theta},\boldsymbol{\rho}\Big)\!\Big), \ \  \forall k \!\in \!\mathcal{K},
\end{equation}
where $B$ is the system bandwidth.

Regarding local computing, the computation rate is $f_{k}/C_{k}$, where $C_k$ is the number of required CPU cycles for users to compute 1-bit of input data, and $f_k$ is the user's CPU frequency. We adopt the dynamic voltage and frequency scaling technique for increasing the computation energy efficiency for users through adaptively controlling the CPU frequency for local computing. Specifically, the energy consumption for the user can be calculated as $L\kappa_kf_k^2$ with $\kappa_k$ being the effective capacitance coefficient of user $k$. Also, as $(1-a_k)E_k$ represents the energy for local computing, we have $(1-a_k)E_k=L\kappa_kf_k^2$. By expressing $f_k$ in terms of other parameters in this equation, we finally obtain the local computation rate of user $k$ as
\begin{equation}
R_{k}^{\mathrm{loc}}\left(a_{k}\right)=\frac{f_{k} }{C_{k}}=\frac{1}{C_{k}} \sqrt{\frac{(1-a_{k})E_{k}}{L\kappa_k}}, \quad \forall k \in \mathcal{K}.
\end{equation}

\subsection{Problem Formulation}

We aim to maximize the computation rate (both due to offloading and local computation) of all the users in a given time slot of duration $L$ through jointly optimizing the phase shifts in $\boldsymbol{\Theta}$, the reflection and transmission amplitude coefficients in $\boldsymbol{\rho}=[\rho_{1}^{\mathrm{t}},\cdots,\rho_{M}^{\mathrm{t}},\rho_{1}^{\mathrm{r}},\cdots,\rho_{M}^{\mathrm{r}}]^{\mathrm{T}}$, the receive beamforming vectors in $\boldsymbol{V}=\left[\boldsymbol{v}_1, \ldots, \boldsymbol{v}_K\right]$, and the energy partition parameters in $\boldsymbol{a}$. The corresponding optimization problem is formally written as
\begin{subequations}
\label{originorigin}
\begin{align}
\label{origin-obj}
\mathscr{P}_{0}\!:\!\max _{\substack{\boldsymbol{a},\boldsymbol{V},\boldsymbol{\Theta},\boldsymbol{\rho}}}
\quad &\sum_{k=1}^{K}\biggl(R_{k}\Big(\boldsymbol{a}, \boldsymbol{v}_{k},\boldsymbol{\Theta},\boldsymbol{\rho}\Big)+R_{k}^{\mathrm{loc}}\left(a_{k}\right)\biggr),\\
\text { s.t. } \ \  & a_{k} \in[0,1], \quad \forall k \in \mathcal{K}, \\
\label{mc1}
&\left|(\boldsymbol{\Theta})_{m,m}\right|=1,\quad \forall m \in \mathcal{M},\\
& \rho_{m}^{\mathrm{t}}, \rho_{m}^{\mathrm{r}} \in\{0,1\}, \quad \forall m \in \mathcal{M},\\
\label{9e}
&\left(\rho_{m}^{\mathrm{t}}\right)^2+\left(\rho_{m}^{\mathrm{r}}\right)^2=1, \quad \forall m \in \mathcal{M}.
\end{align}
\end{subequations}

However, problem (\ref{originorigin}) is a non-convex optimization problem without a closed-form solution as $\boldsymbol{V}$, $\boldsymbol{a}$, $\boldsymbol{\Theta}$, and $\boldsymbol{\rho}$ are highly coupled in (\ref{origin-obj}). Therefore, we apply the BCD framework to $\mathscr{P}_{0}$ so that each block is handled iteratively. Furthermore, due to the binary nature of reflection and transmission amplitude coefficients $\rho_{m}^{\mathrm{t}}$ and $\rho_{m}^{\mathrm{r}}$, problem (\ref{originorigin}) is a nonlinear mixed-integer problem, and it usually requires an exponential time complexity to find the optimal solution. To address the above issue, we apply the logarithmic smoothing-based method to solve the binary variables of $\mathscr{P}_{0}$ in the next section.

\begin{remark}
For a MIMO channel without the STAR-RIS, the number of data streams cannot exceed the degree of freedom (DoF) the channel provides, which is limited by the number of transceiver antennas \cite{tse2005fundamentals}. However, with the presence of a STAR-RIS, currently, there is no study on the rule for the number of antennas at AP $N$ and the number of STAR-RIS elements $M$ to support $K$ users. In Appendix A, we have shown that for the system model considered in this paper, assuming $N \geq K$,  the necessary condition for the number $M$ required to serve $K$ users simultaneously is $M \geq K-b-j$, where $b$ and $j$ are the ranks of the aggregated direct channel between transmission users and the AP $\boldsymbol{H}_{\mathrm{d,t}} =[\boldsymbol{h}_{\mathrm{d},1} \  \cdots \ \boldsymbol{h}_{\mathrm{d},T}]\in\mathbb{C}^{N\times T}$ and reflection users between the AP $\boldsymbol{H}_{\mathrm{d,r}}=[\boldsymbol{h}_{\mathrm{d},T+1} \ \cdots \ \boldsymbol{h}_{\mathrm{d},K}]\in\mathbb{C}^{N\times R}$, respectively. If there are no direct links between transmission and reflection users and the AP, $b$ and $j$ equal zero, and the condition becomes $M \geq K$.
\end{remark}

\section{Logarithmic Smoothing-Based Method for Handling Binary Variables}
\label{3a}
In this section, we optimize $\boldsymbol{\rho}$ with other variables in $\mathscr{P}_{0}$ fixed. We first transform the discontinued binary constraints into their equivalent continuous form and then resort to the penalty-based method. However, it is not likely  to obtain a solution of good quality with the assistance of penalty terms since many undesired stationary points and local optima could be introduced\cite{8378001,murray2010algorithm}, leading to a poor quality converged solution. To diminish the performance loss, we further propose the logarithmic smoothing-based method that eliminates the undesired stationary points and local optima by suppressing the unsmooth part of the penalized problem.

\subsection{Penalty-Based Method for Updating $\boldsymbol{\rho}$}

When other variables are fixed, the subproblem for optimizing $\boldsymbol{\rho}$ is given as
\begin{subequations}
\label{original-rho}
\begin{align}
\max _{\boldsymbol{\rho}} \quad & \sum_{t}R_{t}(\boldsymbol{\rho})+\sum_{r}R_{r}(\boldsymbol{\rho}), \\
\label{bi-sub}
\text { s.t. } \quad & \rho_{m}^{\mathrm{t}}, \rho_{m}^{\mathrm{r}} \in\{0,1\},\quad \forall m \in \mathcal{M},\\
\label{14c}
&\rho_{m}^{\mathrm{t}}+\rho_{m}^{\mathrm{r}}=1,\quad \forall m \in \mathcal{M},
\end{align}
\end{subequations}
where we replaced (\ref{9e}) with (\ref{14c}) since they are equivalent when $\rho_{m}^{\mathrm{t}}, \rho_{m}^{\mathrm{r}} \in\{0,1\}$, and $R_k(\boldsymbol{\rho})=B\log _{2}\Big(1+\gamma_{k}\left(\boldsymbol{\rho}\right)\Big), k \in \{t,r\} $ is obtained from (\ref{rkrho}) by fixing all variables except $\boldsymbol{\rho}$. The binary constraint (\ref{bi-sub}) can be equivalently transformed into the following constraints\cite{borchardt1988exact}
\begin{align}
\label{binary-qcqp}
&\rho_{m}^{\mathrm{x}}-\left(\rho_{m}^{\mathrm{x}}\right)^{2}=0, \quad \forall \mathrm{x} \in\{\mathrm{t}, \mathrm{r}\}, m \in \mathcal{M},\\
\label{0-1}
&0 \leq \rho_{m}^{\mathrm{t}}, \rho_{m}^{\mathrm{r}} \leq 1, \quad m \in \mathcal{M}.
\end{align}
Based on (\ref{0-1}), we always have $\rho_{m}^{\mathrm{x}}-\left(\rho_{m}^{\mathrm{x}}\right)^{2}\geq0$, where the equality holds if and only if $\rho_{m}^{\mathrm{x}}$ is $0$ or $1$, i.e., a binary variable, and the term $\rho_{m}^{\mathrm{x}}-\left(\rho_{m}^{\mathrm{x}}\right)^{2}$ attains its maximum at $\rho_{m}^{\mathrm{x}}=\frac{1}{2}$. Rather than directly handling the nonlinear constraints (\ref{binary-qcqp}), we add a penalty term $\rho_{m}^{\mathrm{x}}-\left(\rho_{m}^{\mathrm{x}}\right)^{2}$ with a penalty parameter $\gamma>0$. The problem (\ref{original-rho}) then becomes
\begin{subequations}
\label{penalty-binary}
\begin{align}
\label{penalty-binary-obj}
\max _{\boldsymbol{\rho}} \ \   & \sum_{t}\!R_{t}(\boldsymbol{\rho})\!+\!\!\sum_{r}\!R_{r}(\boldsymbol{\rho})\!-\!\gamma \!\!\sum_m\!\sum_{\mathrm{x}}\!\!\left(\!\rho_{m}^{\mathrm{x}}-\left(\rho_{m}^{\mathrm{x}}\right)^{2}\right), \\
\label{penalty-a}
\text { s.t. } \ \   & \rho_{m}^{\mathrm{t}}+\rho_{m}^{\mathrm{r}}=1,\quad \forall m \in \mathcal{M},\\
\label{penalty-b}
&0 \leq \rho_{m}^{\mathrm{t}}, \rho_{m}^{\mathrm{r}} \leq 1,\quad \forall m \in \mathcal{M},
\end{align}
\end{subequations}
where the penalty function introduced this way is “exact” in the sense that the problem (\ref{penalty-binary}) and (\ref{original-rho}) have the same global optimum for a sufficiently large penalty parameter $\gamma$. In terms of the penalty parameter, we employ an increasing penalty strategy to enforce the exact constraints successively.

The above idea of applying a penalty term to enforce binary constraint generally performs well, as verified by the simulations in \cite{9570143,9915474}. However, the penalty term might introduce many undesired stationary points and local optima. To avoid getting stuck in undesired stationary points or local optima during the optimization process, we further consider a smoothing technique.

\subsection{Smoothing-Based Optimization Method for Updating $\boldsymbol{\rho}$}

A popular method to eliminate poor local optima is using smoothing methods to suppress the unsmoothness of the objective function \cite{horst2013global}. It aims to transform the original problem with many local optima into one with fewer local optima and thus has a higher chance of obtaining the global optimal solution.

The basic idea of smoothing is to add a strictly convex function (or concave, depending on maximizing or minimizing objective function) to the original objective, i.e.,
\begin{equation}
\label{smooth-eg}
\max_{\boldsymbol{x}, \mu} \ \mathcal{F}(\boldsymbol{x}, \mu)=f(\boldsymbol{x})-\mu \Phi(\boldsymbol{x}),
\end{equation}
where 
$f(\boldsymbol{x})$ is the original objective function,
$\Phi(\boldsymbol{x})$ is a strictly convex function, and $\mu$ is the barrier parameter. If $\Phi(\boldsymbol{x})$ is chosen to have a Hessian that is sufficiently positive definite for all $\boldsymbol{x}$, then $\mathcal{F}(\boldsymbol{x}, \mu)$ is strictly convex when $\mu$ is large enough. This is important in smoothing-based methods because the basic idea is to solve the problem (\ref{smooth-eg}) iteratively for a decreasing sequence of $\mu$ starting with a large value. With a sufficiently large parameter $\mu$ at the beginning, any local optima of $\mathcal{F}(\boldsymbol{x}, \mu)$ is also the unique global optimum and thus is trivial to optimize.

For a binary vector, a strongly convex function $\Phi(\boldsymbol{x})$ is \cite{fiacco1990nonlinear}
\begin{equation}
\label{smooth}
\Phi(\boldsymbol{x})=-\sum_{c=1}^{n}\ln x_c-\sum_{c=1}^{n}\ln (1-x_c).
\end{equation}
This function is well-defined when $ \boldsymbol{0} < \boldsymbol{x} < \boldsymbol{1}$ and attains its maximum at $x_c=\frac{1}{2}$.
Based on (\ref{smooth-eg}) and (\ref{smooth}), we can derive a smoothing-based algorithm for handling (\ref{penalty-binary}). Applying (\ref{smooth}) to the optimization problem (\ref{penalty-binary}), it can be formulated as 
\begin{subequations}
\label{smoothed}
\begin{align}
\label{obj-smoothed}
\begin{split}
\max _{\boldsymbol{\rho}} \  &\sum_{t}\!R_{t}(\boldsymbol{\rho})\!+\!\!\sum_{r}\!R_{r}(\boldsymbol{\rho})\!-\!\gamma \!\!\sum_m\!\sum_{\mathrm{x}}\!\left(\!\rho_{m}^{\mathrm{x}}\!-\!\left(\rho_{m}^{\mathrm{x}}\right)^{2}\!\right)\!-\!\mu\Phi(\boldsymbol{\rho}),
\end{split}\\
\label{non-compact}
\text { s.t. }  \ &\rho_{m}^{\mathrm{t}}+\rho_{m}^{\mathrm{r}}=1,\quad \forall m \in \mathcal{M},\\
\label{0-1v2}
&0 \leq \rho_{m}^{\mathrm{t}}, \rho_{m}^{\mathrm{r}} \leq 1, \quad  m \in \mathcal{M}.
\end{align}
\end{subequations}
Since (\ref{smoothed}) has a continuously differentiable objective function and a  convex feasible set, the projected-gradient (PG) method can be exploited to tackle this problem. It alternatively performs an unconstrained gradient descent step and computes the projection of the update onto the feasible set of the optimization problem. To be specific, the update of the $i^{th}$ iteration is given by
\begin{equation}
\label{gradient-rho}
\boldsymbol{\rho}\left(i+\frac{1}{2}\right)=\boldsymbol{\rho}(i)+\tau(i)\nabla_{\boldsymbol{\rho}} \mathcal{Q}\left(\boldsymbol{\rho}\right),
\end{equation}
where $\mathcal{Q}\left(\boldsymbol{\rho}\right)$ is the objective function in (\ref{smoothed}), $\tau(i)$ is the Armijo step size to guarantee convergence, and $\nabla_{\boldsymbol{\rho}} \mathcal{Q}\left(\boldsymbol{\rho}\right)$ is the gradient of $\mathcal{Q}\left(\boldsymbol{\rho}\right)$, with its explicit expression shown in (\ref{gradient_amp}) of Appendix B.

On the other hand, to project $\boldsymbol{\rho}\left(i+\frac{1}{2}\right)$ onto the feasible set determined by constraints (\ref{non-compact}) and (\ref{0-1v2}), an update point $\boldsymbol{\rho}(i+1)$ can be obtained by solving the following optimization problem
\begin{subequations}
\label{proj}
\begin{align}
\boldsymbol{\rho}(i+1)=\argmin_{\boldsymbol{\rho}}  \quad &\left\|\boldsymbol{\rho}-\boldsymbol{\rho}\left(i+\frac{1}{2}\right)\right\|^{2},\\
\label{matrix-compact}
\text { s.t. } \quad & \left[\begin{array}{cc}
\!\!I_{M}\!\!\!\!&I_{M}\!\!\!\! \\
\end{array}\right] \boldsymbol{\rho} = \boldsymbol{1},\\
\quad & \ \boldsymbol{0} \leq \boldsymbol{\rho} \leq \boldsymbol{1},
\end{align}
\end{subequations}
where $\left[\begin{array}{cc}
\!\!I_{M}&\!\!\!\!I_{M}\!\!\! \\
\end{array}\right] \in \mathbb{R}^{M \times 2M}$ and we transformed the constraints (\ref{non-compact}) into its equivalent compact matrix form (\ref{matrix-compact}). Since the feasible set of $\boldsymbol{\rho}$ is the intersection of a hyperplane and a box, a closed-form solution can be derived based on the Karush-Kuhn-Tucker (KKT) condition and is given by the following Proposition 1, which is proved in Appendix C. Based on (\ref{gradient-rho}), (\ref{gradient_amp}) and (\ref{proj-1}), we can iteratively update $\boldsymbol{\rho}$, where the convergent point is guaranteed to be a stationary point of (\ref{smoothed}).
\begin{pro1}
The optimal solution to (\ref{proj}) is given by
\begin{equation}
\label{proj-1}   \boldsymbol{\rho}^{\star}=P_{\operatorname{Box}[\boldsymbol{0}, \boldsymbol{1}]}\left(\boldsymbol{\rho}\left(i+\frac{1}{2}\right)-\begin{bmatrix}
\boldsymbol{\lambda}^{\star} \\
\boldsymbol{\lambda}^{\star}
\end{bmatrix}\right),
\end{equation}
where $\boldsymbol{\lambda}^{\star}=[\lambda_1^{\star}, \ldots, \lambda_K^{\star}]$ and $P_{\operatorname{Box}[\boldsymbol{0}, \boldsymbol{1}]}(\boldsymbol{x})=\left(\min \left\{\max \left\{x_{i}, 0\right\}, 1\right\}\right)_{i=1}^{n}, \boldsymbol{x} \in \mathbb{R}^{n}$. The $\lambda_m^{\star}$ is a solution to the equation
\begin{equation}
\label{proj-2}
\left[\begin{array}{cc}
\!\!I_{M}&\!\!\!\!I_{M}\!\!\! \\
\end{array}\right]P_{\operatorname{Box}[\boldsymbol{0}, \boldsymbol{1}]}\biggl(\boldsymbol{\rho}\left(i+\frac{1}{2}\right)-\begin{bmatrix}
\boldsymbol{\lambda}^{\star} \\
\boldsymbol{\lambda}^{\star}
\end{bmatrix}\biggl)=\boldsymbol{1},
\end{equation}
and can be obtained via the bisection search.
\end{pro1}

In order to enforce the smoothness induced by the last term of (\ref{obj-smoothed}), the problem (\ref{smoothed}) is solved for a sequence of decreasing values of $\mu$ since the solution to (\ref{smoothed}), $\boldsymbol{\rho}^{\star}(\mu)$, is a continuously differentiable function \cite{bertsekas1997nonlinear}. Initially, it starts with a suitably large $\mu=\mu_0$ since it is vital that the iterates move away from an undesired local optimum. After obtaining the solution of (\ref{smoothed}) with $\mu=\mu_0$, it is used as the starting point of solving (\ref{smoothed}) but with $\mu=\mu_1=\eta_\mu \mu_0$ where $\eta_\mu <1$. The iteration goes on until $\mu_k$ reaches the tolerance for barrier value $\epsilon_{\mu}$\cite{murray2002algorithms}. In terms of $\gamma$, it can be changed synchronously with $\mu$ by a multiplicative factor $\eta_{\gamma}>1$ since both smoothing-based and penalty-based methods use solution $\boldsymbol{\rho}$ from the previous iteration as the starting point of the next iteration. The overall algorithm for solving $\boldsymbol{\rho}$ under the proposed logarithmic smoothing framework is summarized in Algorithm \ref{rho-sm}. Since the barrier parameter ends with a value that is close to zero\cite{murray2010algorithm}, for a sufficiently large value of the penalty parameter $\gamma$, Algorithm \ref{rho-sm} will obtain a stationary point of the original problem (\ref{original-rho})\cite{9570143}. The performance improvement of Algorithm \ref{rho-sm} compared with the conventional penalty-based method will be illustrated later in Section \ref{5a}.


\RestyleAlgo{ruled}
\begin{algorithm}[t]
    \SetKwInOut{Input}{Input}
    \SetKwInOut{Output}{Output}
    \SetKwInOut{Set}{Set}
\caption{Logarithmic Smoothing-based Method for Solving (\ref{original-rho})}\label{rho-sm}
\Input{$\epsilon_{F}, \epsilon_{\mu}, \epsilon_{\gamma}, \eta_{\mu}, \eta_{\gamma}, \mu_{0}$ and $\gamma_{0} $.}
\Set{$\mu=\mu_0$ and $\gamma=\gamma_0$.}
\While{$ \gamma < \epsilon_{\gamma}$ \text{or} $\mu > \epsilon_{\mu}$}{
\While{The increase of the objective value of (\ref{smoothed}) is above $\epsilon_{F}$}
{$\begin{aligned} &\text { Update } \boldsymbol{\rho} \text{ based on equation (\ref{gradient-rho}) and (\ref{proj-1})}.\end{aligned}$}

$\begin{aligned}  &\text { Update the penalty parameter } \gamma \leftarrow \eta_{\gamma} \gamma . \\  &\text { Update the barrier parameter } \mu \leftarrow \eta_{\mu} \mu . \end{aligned}$}
\end{algorithm}

\section{Fast Iterative BCD Algorithm for Solving $\mathscr{P}_0$}
\label{3}

With the binary transmission and reflection amplitude coefficients solved in Section \ref{3a}, this section derives the details of optimization algorithms for other subproblems under the BCD framework. Specifically, to facilitate a fast iterative algorithm, all the subproblems are solved in closed-form or with a first-order algorithm, with complexity orders only scaling linearly with respect to the number of elements in the STAR-RIS or the number of users. 

\subsection{Updating $\boldsymbol{a}$}

When other variables are fixed, the subproblem for updating $\boldsymbol{a}$ is
\begin{subequations}
\label{energyp}
\begin{align}
\label{energyp-obj}
\max _{\substack{\boldsymbol{a}}} \quad
&\sum_{k=1}^{K}R_{k}(\boldsymbol{a})+\sum_{k=1}^{K}R_{k}^{\mathrm{loc}}\left(a_{k}\right),\\
\label{0-1con}
\text { s.t. } \quad & a_{k} \in[0,1], \quad \forall k \in \mathcal{K}.
\end{align}
\end{subequations}
Conventionally, since $R_{k}\left(\boldsymbol{a}\right)$ in (\ref{energyp-obj}) can be re-expressed as the difference of two concave functions, the DC programming method can be applied to convexify the problem (\ref{energyp}) and further solved numerically by the existing convex solvers such as CVX (the second term in (\ref{energyp-obj}) is already concave). However, the above method is a multi-stage iterative optimization algorithm with the outer loop involving DC programming, and the inner loop still requires iterative numerical methods, thus incurring high computational complexity\footnote{A common practice is to use the interior-point method, and its complexity order is at least $\mathcal{O}(K^{3.5})$.}. To overcome this difficulty, we propose a new approach based on a Lagrangian dual reformulation of the energy partition maximization problem and subsequently apply the quadratic transform method. This leads to an algorithm in which each iteration is performed in closed-forms and bisection search rather than being solved with numerical convex solvers. Thus the proposed method is more desirable and computationally efficient.

First, to establish the legitimacy of applying Lagrangian dual reformulation of (\ref{energyp}), we provide the following property, which is proved in Appendix D.
\begin{pro1}
\label{prop1}
The energy partition maximization problem (\ref{energyp}) is equivalent to
\begin{equation}
\label{dta-obj}
\begin{aligned}
\max _{\boldsymbol{a},\{\eta_k\}_{k=1}^{K}} \quad
&\mathcal{D}\left(\boldsymbol{a},\{\eta_k\}_{k=1}^{K}\right),\\
\text { s.t. } \quad & a_{k} \in[0,1], \quad \forall k \in \mathcal{K},
\end{aligned}
\end{equation}
where $\eta_k$ refers to the auxiliary variable, and the new objective is given by
\begin{equation}
\label{DD}
\begin{aligned}
\mathcal{D} &\left(\boldsymbol{a},\{\eta_k\}_{k=1}^{K}\right)=\sum_{k \in \mathcal{K}} \log_2 (\left.1+\eta_k\right)-\sum_{k \in \mathcal{K}} \frac{\eta_k}{\ln 2} \\
+& \frac{1}{\ln 2} \sum_{k \in \mathcal{K}} \frac{\left(1+\eta_k\right)\left|\left(\boldsymbol{v}_{k}\right)^{\mathrm{H}}\!
\boldsymbol{g}_{k}
\right|^{2} a_{k} {E}_{k}}{\sum_{l \in \mathcal{K}}\left|\left(\boldsymbol{v}_{k}\right)^{\mathrm{H}}\!
\boldsymbol{g}_{l}
\right|^{2} a_{l} {E}_{l}
+L\sigma^2\left\|\boldsymbol{v}_k\right\|^2}\\ +&\sum_{k \in \mathcal{K}}\frac{1}{C_{k}} \sqrt{\frac{(1-a_{k})E_{k}}{L\kappa_k}}.
\end{aligned}
\end{equation}
\end{pro1}

Based on Proposition 2, variables $\boldsymbol{a}$ and $\{\eta_k\}_{k=1}^{K}$ are updated iteratively. When $\boldsymbol{a}$ is fixed, the function $\mathcal{D}$ is concave with respect to $\eta_k$. Therefore, the global optimal $\eta_k$ is obtained by setting $\partial \mathcal{D} / \partial \eta_k$ to zero, i.e.,
\begin{equation}
\label{updateeat}
\eta_k^{\star} = \frac{a_{k} {E}_{k}\left|\left(\boldsymbol{v}_{k}\right)^{\mathrm{H}}\!
\boldsymbol{g}_{k}
\right|^{2}}{\sum_{l \neq k} a_{l} {E}_{l}\left|\left(\boldsymbol{v}_{k}\right)^{\mathrm{H}} \boldsymbol{g}_{l}
\right|^{2}\!+\!L\sigma^{2}\left\|\boldsymbol{v}_{k}\right\|^{2}}, \quad \forall k \in \mathcal{K}.
\end{equation}

To optimize {$\boldsymbol{a}$ when $\{\eta_k\}_{k=1}^{K}$ is fixed, we apply the quadratic transform on the fractional term in (\ref{DD}). In particular, when $\eta_k$ is held fixed, only the last two terms of $\mathcal{D}$, which are a sum-of-ratio form and a sum of square roots, are related to the optimization of $a_k$. In general, if the objective function is a sum of fractional forms with the fraction’s numerator being concave and the denominator being convex (known as the concave-convex form), one can employ the quadratic transform to convert the concave-convex fractional programming into a sequence of convex subproblems\cite{8314727,8310563}. Since the sum of square roots can be considered as a special case of sum-of-ratio form with a denominator equal to 1, by the quadratic transform, we further recast $\mathcal{D}$ to
\begin{equation}
\label{md}
\begin{aligned}
&\mathcal{D}_{q}\left(\boldsymbol{a},\{y_k,z_k\}_{k=1}^{K}\!\right)\!=\!\sum_{k \in\mathcal{K}}\!\!\left(\!\!\sqrt{\frac{4z_k^2}{C_k}}\sqrt[4]{\frac{(1-a_{k})E_{k}}{L\kappa_k}}\!-\!z_k^2\right)\!\!\\
&+\!\!\sum_{k \in \mathcal{K}}\!2y_k \sqrt{\!\left(1+\eta_k\right)\!\left|\left(\boldsymbol{v}_{k}\right)^{\mathrm{H}}\!
\boldsymbol{g}_{k}
\right|^{2}\!\!a_{k}{E}_{k}}\!+\!\!\sum_{k \in \mathcal{K}}\!\log_2 (\!\left.1\!+\!\eta_k\right) \\
&- \!\! \sum_{k \in \mathcal{K}} y_k^2\left(\sum_{l \in \mathcal{K}}\left|\left(\boldsymbol{v}_{k}\right)^{\mathrm{H}}\!\boldsymbol{g}_{l}\right|^{2}\!\!a_{l} {E}_{l}\! +\! L\sigma^2\left\|\boldsymbol{v}_k\right\|^2\right)\!\!-\sum_{k \in \mathcal{K}} \eta_k,
\end{aligned}
\end{equation}
where $\{y_k,z_k\}_{k=1}^{K}$ are the auxiliary variables. Then according to the iterative update rules of the quadratic transform, with fixed $\boldsymbol{a}$, the optimal $\{y_k,z_k\}_{k=1}^{K}$ can be directly given as
\begin{equation}
\label{updatez}
y_k^{\star}=\frac{\sqrt{\left(1+\eta_k\right)\left|\left(\boldsymbol{v}_{k}\right)^{\mathrm{H}}\!
\boldsymbol{g}_{k}
\right|^{2} a_{k}{E}_{k}}}{\sum_{l \in\mathcal{K}}\left|\left(\boldsymbol{v}_{k}\right)^{\mathrm{H}}\!\boldsymbol{g}_{l}\right|^{2} a_{l} {E}_{l} + L\sigma^2\left\|\boldsymbol{v}_k\right\|^2}, \quad \forall k \in \mathcal{K},
\end{equation}
\begin{equation}
\label{updatey}
z_k^{\star}=\sqrt[4]{\frac{(1-a_{k})E_{k}}{L\kappa_k}}, \quad \forall k \in \mathcal{K}.
\end{equation}
On the other hand, since $\mathcal{D}_{q}$ is concave with respect to $\boldsymbol{a}$ with fixed $\{y_k,z_k\}_{k=1}^{K}$, the iterative update for $a_k$ is obtained by setting $\partial \mathcal{D}_{q}/ \partial a_k$ to zero, i.e.,
\begin{equation}
\label{syy}
    s_k(1-a_k)^{-\frac{3}{4}}+o_k(a_k)^{-\frac{1}{2}}-w_k = 0, \ a_k \in [0,1],
\end{equation}
where 
\begin{subequations}
    \begin{align}
        w_k&=\left(\sum_{k \in \mathcal{K}}y_k^2\right)|(\boldsymbol{v}_{k})^{\mathrm{H}}\boldsymbol{g}_{k}|^{2}{E}_{k}>0,\\ o_k&=y_k\sqrt{(1+\eta_k)\left|\left(\boldsymbol{v}_{k}\right)^{\mathrm{H}}\!\boldsymbol{g}_{k}\right|^{2}{E}_{k}}>0,\\
        s_k &= -\sqrt{\frac{z_k^2}{4C_k}}(\frac{E_k}{L\kappa_k})^{\frac{1}{4}}<0.
    \end{align}
\end{subequations}
Note that equation (\ref{syy}) is a strictly decreasing function with respect to $a_k \in [0,1]$ since its derivative is derived as $\frac{3}{4}s_k(1-a_k)^{-\frac{7}{4}}-\frac{o_k}{2}a^{-\frac{3}{2}}<0$. Furthermore, if $a_k$ is 0 or 1, the left-hand side of the equation (\ref{syy}) will go to infinity and minus infinity, respectively. The above characteristics imply that equation (\ref{syy}) must have a unique solution, and a bisection search algorithm can be used to find the optimal $a_k^{\star}$. The details are omitted here for brevity.



These updating steps amount to an iterative optimization as stated in Algorithm \ref{energy-cf}. The objective function (\ref{DD}) is bounded above and monotonically nondecreasing after each iteration. The solution of Algorithm \ref{energy-cf} arrives at a stationary point of the reformulated problem (\ref{dta-obj}). Since the problem (\ref{dta-obj}) is equivalent to (\ref{energyp}), it is a stationary point of the original problem (\ref{energyp})\cite{8310563}.


\RestyleAlgo{ruled}
\begin{algorithm}[t]
    \SetKwInOut{Input}{Input}
    \SetKwInOut{Output}{Output}
    \SetKwInOut{Set}{Set}
\caption{Proposed Algorithm for Energy Partition Maximization (\ref{energyp}) }\label{energy-cf}
\Input{Initial $a_k$ and $\eta_k$.}
\While{Stopping criterion is not satisfied}{
$\begin{aligned}  \text { Update } \eta_k \text{ by equation (\ref{updateeat}).}\end{aligned}$\\
\While{Stopping criterion is not satisfied}{$\begin{aligned}  &\text { Update } y_k \text{ and } z_k \text{ by equation (\ref{updatey}) and (\ref{updatez}).}\\
&\text{ Update } \!a_k\! \text{ by applying bisection search method} \\  &\text{ on (\ref{syy})}.
\end{aligned}$}}
\end{algorithm}
\subsection{Updating $\boldsymbol{V}$}
When other variables are fixed, the subproblem of $\mathscr{P}_{0}$ for updating $\boldsymbol{V}$ is
\begin{equation}
\label{vk}
\max _{\substack{\boldsymbol{V}}} \quad \sum_{k=1}^{K}R_{k}\left(\boldsymbol{v}_{k}\right).
\end{equation}
Since the uplink SINR of user $k$ only contains its own receive beamforming vector $\boldsymbol{v}_k$, we can optimize $\boldsymbol{v}_k$ in a parallel manner, and the objective function is simply the SINR in (\ref{upsinr}). This gives the $k^{th}$ subproblem as
$
\max _{\boldsymbol{v}_{k}} \gamma_{k}\left(\boldsymbol{v}_{k}\right)=\frac{\boldsymbol{v}_{k}^{\mathrm{H}} \boldsymbol{A}_{k} \boldsymbol{v}_{k}}{\boldsymbol{v}_{k}^{\mathrm{H}} \boldsymbol{B}_{k} \boldsymbol{v}_{k}},
$
where $\boldsymbol{A}_{k}=p_{k}\boldsymbol{g}_{k}\left(\boldsymbol{g}_{k}\right)^{\mathrm{H}}$ and $\boldsymbol{B}_{k}=\sum_{i=1, i \neq k}^{K} p_{i} \boldsymbol{g}_{i}\left(\boldsymbol{g}_{i}\right)^{\mathrm{H}}+$ $\sigma^{2} \mathbf{I}_{N}$. It is easy to recognize that this is a generalized eigenvector problem, and its optimal solution $\boldsymbol{v}_{k}^{\star}$ should be the eigenvector corresponding to the largest eigenvalue of the matrix $\left(\boldsymbol{B}_{k}\right)^{-1} \boldsymbol{A}_{k}$. 

\subsection{Updating $\boldsymbol{\Theta}$}

When other variables are fixed, the subproblem for updating STAR-RIS phase shifts is given by
\begin{subequations}
\label{xi-grad}
\begin{align}
\max _{\boldsymbol{\Theta}} \quad
&\sum_{t}R_{t}\left(\boldsymbol{\Theta}\right)+\sum_{r}R_{r}\left(\boldsymbol{\Theta}\right),\\
\label{modd}
\text { s.t. } \quad &\left|(\boldsymbol{\Theta})_{m,m}\right|=1,\quad \forall m \in \mathcal{M}.
\end{align}
\end{subequations}
To handle the modulus constraints in (\ref{modd}), the semi-definite relaxation (SDR) method and manifold method are widely adopted. However, SDR-based methods only yield an approximate solution without an optimality guarantee and incur a heavy computational burden as the variable's dimension increases. Furthermore, manifold optimization methods slow down the convergence due to their nested loop architecture\cite{9367432}. Therefore, we propose a  gradient descent (GD) method algorithm as follows. In contrast to the SDR and manifold optimization-based methods, this method avoids convex relaxation, does not need to solve high-dimensional SDP problems, and gets rid of the nested loops. Thus, it significantly improves computational efficiency.

Observe that the unknown variable $\boldsymbol{\Theta}$ in the feasible set (\ref{modd}) is in fact $\left\{\theta_{m}\right\}_{m=1}^{M}\in[0,2 \pi) $. Therefore, problem (\ref{xi-grad}) can be recast into an unconstrained optimization problem as
\begin{equation}
\label{theta-grad}
\max _{\{\theta_m\}_{m=1}^{M}} \!\!\!\!\mathcal{H}\!\left(\!\{\theta_m\}_{m=1}^{M}\!\right)\!\!:=\!\!\sum_{t}\!R_{t}\!\left(\!\{\theta_m\!\}_{m=1}^{M}\!\right)\!+\!\sum_{r}\!R_{r}\!\left(\!\{\theta_m\}_{m=1}^{M}\!\right)\!,
\end{equation}
where we drop the constraints $\theta_m \in[0,2 \pi), \forall m \in \mathcal{M}$ since the objective function is periodic and taking modulus of $2\pi$ can recover a $\theta_m \in[0,2 \pi)$.
Since (\ref{theta-grad}) is an unconstrained optimization problem with a differentiable objective function, it can be readily solved by the GD method to obtain a stationary point \cite{lipp2016variations}. To be specific, the update of $\theta_{m}$ at the $i^{th}$ iteration is given by
\begin{equation}
\label{gradient-theta}
\theta_m\left(i+1\right)=\theta_m(i)+\tau(i)\nabla_{\theta_m} \mathcal{H}\left(\{\theta_m\}_{m=1}^{M}\right),
\end{equation}
where $\tau(i)$ is the Armijo step size to guarantee convergence and $\nabla_{\theta_m} \mathcal{H}\left(\{\theta_m\}_{m=1}^{M}\right)$ is the gradient of $\mathcal{H}\left(\{\theta_m\}_{m=1}^{M}\right)$ given by
\begin{equation}
\label{gradient_rho}
\begin{aligned}
\nabla&\!_{\theta_m} \!\mathcal{H}\!\left(\theta_m\right)\!\!=\!\!\!\sum_k\!\!\Bigg(\!\!\frac{1}{\ln\!2(\!1\!+\!\gamma_{k}\!)}\!2\!\operatorname{Re}\!\!\left(\!\!\frac{\!\left(\!p_{k}\boldsymbol{G}^{*}\boldsymbol{v}_{k}^{*}\boldsymbol{v}_{k}^{\mathrm{T}}\boldsymbol{g}_{k}^{*}\boldsymbol{h}_{\mathrm{r}, k}^{\mathrm{T}}\!\right)_{m,m}\!\!je^{j\theta_m}}{\sum_{l \neq k} p_{l}\!\left|\left(\boldsymbol{v}_{k}\right)^{\mathrm{H}}\!\boldsymbol{g}_{l}
\right|^{2}\!\!\!+\!\sigma^{2}\!\left\|\boldsymbol{v}_{k}\right\|^{2}}\right. \\&-\left. \left.\frac{p_{k}\left|\left(\boldsymbol{v}_{k}\right)^{\mathrm{H}}\!
\boldsymbol{g}_{k}
\right|^{2}\sum_{l \neq k}\left(p_{l}\boldsymbol{G}^{*}\boldsymbol{v}_{k}^{*}\boldsymbol{v}_{k}^{\mathrm{T}}\boldsymbol{g}_{l}^{*}\boldsymbol{h}_{\mathrm{r}, l}^{\mathrm{T}}\right)_{m,m}je^{j\theta_m}}{\left(\sum_{l \neq k} p_{l}\left|\left(\boldsymbol{v}_{k}\right)^{\mathrm{H}} \boldsymbol{g}_{l}
\right|^{2}\!+\!\sigma^{2}\left\|\boldsymbol{v}_{k}\right\|^{2}\right)^2}\right)\!\!\right).
\end{aligned}
\end{equation}

\subsection{Summary and Computation Complexity}
The proposed four-block BCD optimization algorithm for solving the computation rate maximization problem $\mathscr{P}_{0}$ in (\ref{originorigin}) is summarized in Algorithm \ref{all-bf}, where the converged point is guaranteed to a stationary point \cite{razaviyayn2013unified}.

\RestyleAlgo{ruled}
\begin{algorithm}[t]
    \SetKwInOut{Input}{Input}
    \SetKwInOut{Output}{Output}
\caption{Overall Algorithm for Solving Computation Rate Maximization Problem $\mathcal{P}_0$}\label{all-bf}
\Input{$K,M,L,B,N,\{E_{k}, C_{k}, \kappa_{k}, \boldsymbol{h}_{\mathrm{d}, k}, \boldsymbol{h}_{\mathrm{s}, k}\}_{k \in \mathcal{K}}$, $\boldsymbol{G}$.}
\While{Stopping criterion is not satisfied}{

$\begin{aligned} &\text { Update } \{\boldsymbol{v}_k\}_{k=1}^{K} \text{ as the eigenvector for } \boldsymbol{B}_{k}^{-1} \boldsymbol{A}_k \\ 
&\text{ corresponding to its largest eigenvalue}. \\ \end{aligned}$

\While{Stopping criterion is not satisfied}
{$\begin{aligned} \text { Update } \left\{\theta_{m}\right\}_{m=1}^{M} \text{ based on equation (\ref{gradient-theta})}. \end{aligned}$}
$\begin{aligned}  &\text { Solving } \boldsymbol{\rho} \text { by using Algorithm \ref{rho-sm}}. \\
&\text { Solving } \boldsymbol{a} \text { by using Algorithm \ref{energy-cf}}. 
\end{aligned}$}
\end{algorithm}

Notice that the iteration with respect to $\{\theta_m\}_{m=1}^{M}$ is based on the GD method, which only involves first-order differentiation. Therefore, it has $\mathcal{O}(M)$ complexity order. On the other hand, the computational complexity of the iteration with respect to $\boldsymbol{\rho}$ is dominated by the gradient step. Hence, the complexity order for updating $\boldsymbol{\rho}$ is also $\mathcal{O}(M)$. Furthermore, the complexity to update $\boldsymbol{V}$ and $\boldsymbol{a}$ in Algorithm \ref{energy-cf} is $\mathcal{O}(KN^3)$ and $\mathcal{O}(K)$, respectively \cite{8982186}.  Based on the above discussions, the complexity order of solving (\ref{originorigin}) is linear in $M$ and $K$. This makes the proposed algorithm suitable for massive elements and user networks. The computation complexity of the proposed algorithm is summarised in Table \ref{table1}.

\begin{table*}[htbp]
\caption{The computation complexity of the proposed algorithm.}
\label{table1}  
\centering  
\begin{tabular}{|c|c|c|c|c|}
\hline
 & Updating $\boldsymbol{\Theta}$ & Updating $\boldsymbol{\rho}$ & Updating $\boldsymbol{a}$ & Updating $\boldsymbol{V}$ \\
\hline
Computation complexity & $\mathcal{O}(M)$ & $\mathcal{O}(M)$ &$\mathcal{O}(K)$ & $\mathcal{O}(KN^3)$\\
\hline
\end{tabular}
\end{table*}

\section{Simulation Results}
In this section, we present simulation results to verify the effectiveness of the proposed algorithm. As illustrated in Fig. \ref{sl}, under a three-dimensional Cartesian coordinate system, we consider a system with 4 users in transmission space and 4 users in reflection space, and they are uniformly and randomly located in a square region of $50 \mathrm{~m} \times 50 \mathrm{~m}$ centered at the 3-dimensional coordinate $(45,0,0)$ and $(95,0,0)$, respectively. The STAR-RIS and  AP are located at the 3-dimensional coordinate $(75,0,15)$ and $(0,0,15)$, respectively.
\begin{figure}[t]
\centering
\centerline{\includegraphics[scale=0.11]{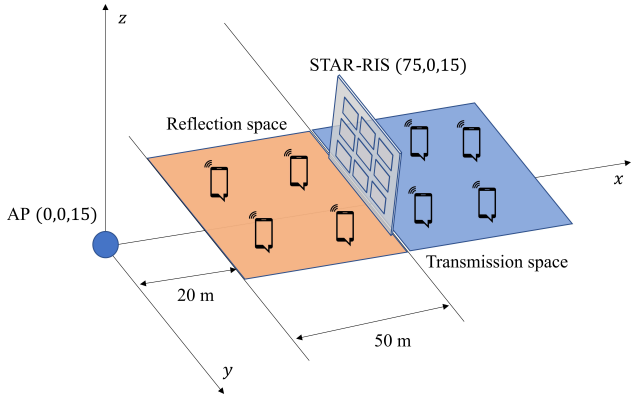}}
\caption{The simulated STAR-RIS-aided MEC network scenario.}
\label{sl}
\end{figure}

Spatially independent Rician fading channel is considered for all channels to account for both the line-of-sight (LoS) and non-LoS (NLoS) components \cite{tse2005fundamentals}. For example, the AP-RIS channel is expressed as $\boldsymbol{G}=\sqrt{L_{\mathrm{AR}}(d)}\left(\sqrt{\frac{\kappa_{\mathrm{AR}}}{1+\kappa_{\mathrm{AR}}}} \boldsymbol{G}^{\mathrm{LoS}}+\sqrt{\frac{1}{1+\kappa_{\mathrm{AR}}}} \boldsymbol{G}^{\mathrm{NLoS}}\right)$, where $\kappa_{\mathrm{AR}}$ is the Rician factor representing the ratio of power between the LoS path and the scattered paths, $\boldsymbol{G}^{\mathrm{LoS}}$ is the LoS component modeled as the product of the unit spatial signature of the AP-RIS link \cite{tse2005fundamentals,9110912}, $\boldsymbol{G}^{\text {NLoS}}$ is the Rayleigh fading components with entries distributed as $\mathcal{C} \mathcal{N}(0,1)$, and $L_{\mathrm{AR}}(d)$ is the distance-dependent path loss of the AP-RIS channel. We consider the following distance-dependent path loss model $L_{\mathrm{AR}}(d)=T_{0}\left(\frac{d}{d_{0}}\right)^{-\alpha_{\mathrm{AR}}}$, where $T_{0}$ is the constant path loss at the reference distance $d_{0}=1 \mathrm{~m}, d$ is the Euclidean distance between the transceivers, and $\alpha_{\mathrm{AR}}$ is the path loss exponent\cite{8811733}. Since the STAR-RIS can be practically deployed in LoS with the AP, we set $\alpha_{\mathrm{AR}}=2$ and $\kappa_{\mathrm{AR}}=30 \mathrm{~dB}$ \cite{8811733}. In addition, other channels are similarly generated with $\alpha_{\mathrm{AU}}=3.5$ and $\kappa_{\mathrm{AU}}=0$ (i.e., Rayleigh fading to account for rich scattering) for the AP-user channel, $\alpha_{\mathrm{RU}}=2.5$ and $\kappa_{\mathrm{RU}}=3$ for the RIS-user channel\cite{9352968,9380744}. We consider a system with a bandwidth $1~\mathrm{MHz}$, $T_{0}=-30 \mathrm{~dB}$ and the effective noise power density is $-150~\mathrm{dBm/Hz}$\cite{8930608}. Thus the noise power at the AP is $\sigma^{2}=-90~\mathrm{dBm}$. Each antenna at the AP is assumed to have an isotropic radiation pattern, and the antenna gain is $0~\mathrm{dBi}$ \cite{9133435}. Unless specified otherwise, other parameters are set as follows: $E_{k}=10 \mathrm{~J}$, $C_k=200 \mathrm{~cycles/bit}$, $\kappa_k = 10^{-25}$, and $L=1 \mathrm{~s}$\cite{9380744}. 

\subsection{Performance of the Proposed Algorithms}
\label{5a}
First, we demonstrate the convergence behavior of Algorithm 1 compared with the penalty-based method in Section \ref{3a} for solving (\ref{original-rho}). Note that both the smoothing-based and penalty-based methods are solved by the PG method for a fair comparison.

The superiority of the proposed smoothing-based method in terms of the objective values is shown in Fig. \ref{bvpconvergence} under a specific channel realization. It is observed that the proposed algorithm achieves a higher computation rate compared with the penalty-based method. This shows the effectiveness of the proposed smoothing-based method in eliminating the poor local optima and therefore achieving a higher computation rate. 

Next, in Fig. \ref{FPconvergence}, we compare the convergence behavior of Algorithm 2 and the DC programming with CVX for solving (\ref{energyp}). It shows that Algorithm 2 and the DC programming reach almost the same objective value at the end. Although DC programming converges in fewer iterations than Algorithm 2, Algorithm 2 is much more efficient than DC programming on a per-iteration basis since Algorithm 2 updates variables in closed-forms or via bisection search, while DC programming requires solving a convex optimization numerically in each iteration. This is reflected by the computation time of Algorithm 2 to reach a computation rate of $4.61\times10^7$ being only 0.057 s while that of the DC programming is 3.116 s.

\begin{figure}[t]
     \centering
     \begin{subfigure}[b]{0.49\textwidth}
     \centering
  \includegraphics[scale=0.5]{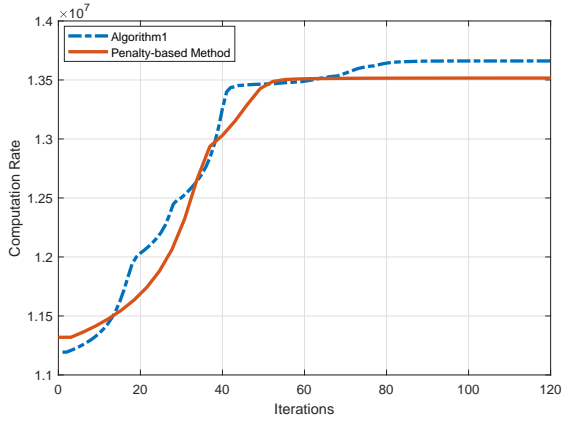}
        \caption{Convergence behavior of the proposed smoothing-based method and penalty-based method.}
        \label{bvpconvergence}
     \end{subfigure}
     \hfil
     \begin{subfigure}[b]{0.49\textwidth}
         \centering
  \includegraphics[scale=0.5]{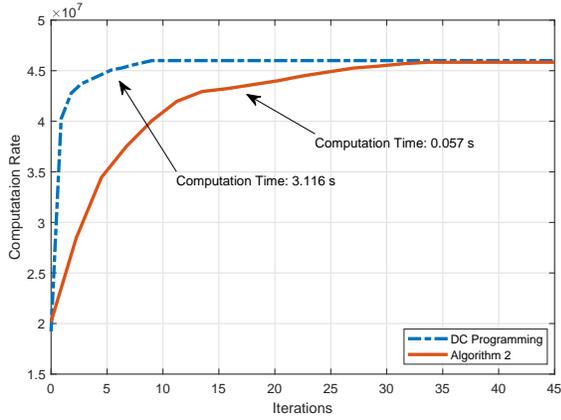}
        \caption{Convergence behavior of the Algorithm 2 and DC programming with CVX.}
        \label{FPconvergence}
     \end{subfigure}
        \caption{Comparisons of convergence behavior with $M=30$, $N = 10$, $K = 8$.}
\end{figure}

In addition, we demonstrate the overall convergence of the proposed algorithm (Algorithm \ref{all-bf}) and compare with the following two benchmarks: SDR-DC method for solving ${\{\theta_m\}_{m=1}^{M}}$ via SDR and solving $\boldsymbol{a}$ with DC programming, and the ‘Penalty-based Method’ for solving (\ref{original-rho}) with the penalty-based method as in Section \ref{3a} with exact binary constraints. 
\begin{figure}[t]
     \centering
  \includegraphics[scale=0.5]{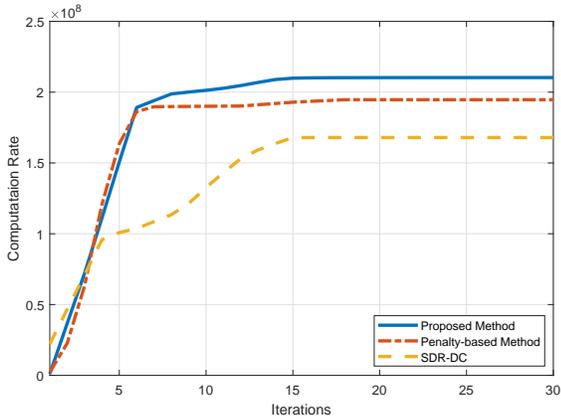}
        \caption{Convergence behaviour of the proposed algorithms with $M=30$, $N = 10$, $K = 8$.}
        \label{convergence}
\end{figure}

The superiority of the proposed algorithm in terms of the computation rate is shown in Fig. \ref{convergence} under a specific channel realization. It is observed that the proposed algorithm achieves a higher computation rate than the penalty-based method. On the other hand, as the proposed algorithm include an extra smoothing parameters update step, it takes more steps in outer iterations in the BCD algorithm to converge. It is also observed that both the proposed and the penalty-based methods outperform the SDR-DC method in convergence speed and computation rate performance. 

\begin{figure}[t]
     \centering
  \includegraphics[scale=0.5]{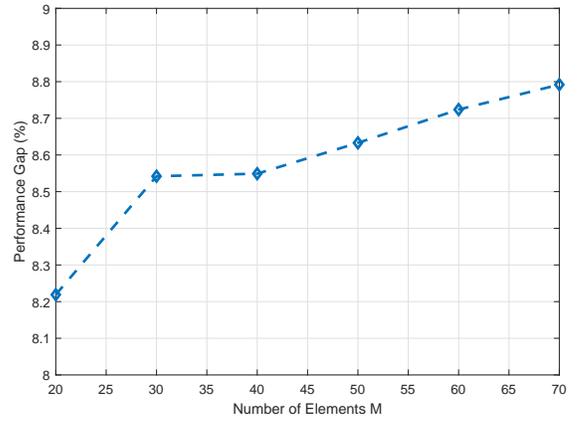}
        \caption{Performance gains over the penalty-based method with $N = 10$, $K = 8$.}
        \label{pg}
\end{figure}

In Fig. \ref{pg}, we show the average performance gap between the proposed and penalty-based methods versus the number of elements in STAR-RIS. We can see that the proposed method outperforms the penalty-based method in terms of the objective values obtained. Moreover, as the number of elements increases, the performance gap widens in general. It illustrates the potential of the proposed smoothing-based method in solving a massive STAR-RIS-aided wireless system.

To further show the low computational complexity of the proposed algorithm, we compare the computation time with SDR-DC and penalty-based methods. As shown in Fig. \ref{CT} and Fig. \ref{CTU}, with the number of elements or users increases, the proposed method and penalty-based method save the computation time to a large extent compared with the SDR-DC method (e.g., more than 10 times differences with 60 elements and 8 times differences with 14 users, in Fig. \ref{CT} and Fig. \ref{CTU}, respectively), and the advantage becomes more prominent as $K$ or $M$ increases. On the other hand, it shows that the proposed algorithm achieves almost the same computation time compared with the penalty-based method for small numbers of $M$ and $K$. As $M$ and $K$ increase, the computation time of the proposed algorithm is only marginally higher than that of the penalty-based method. Both the proposed and penalty-based methods are suitable for large-scale STAR-RIS optimization.
\begin{figure}[t]
     \centering
     \begin{subfigure}[b]{0.49\textwidth}
          \centering
  \includegraphics[scale=0.5]{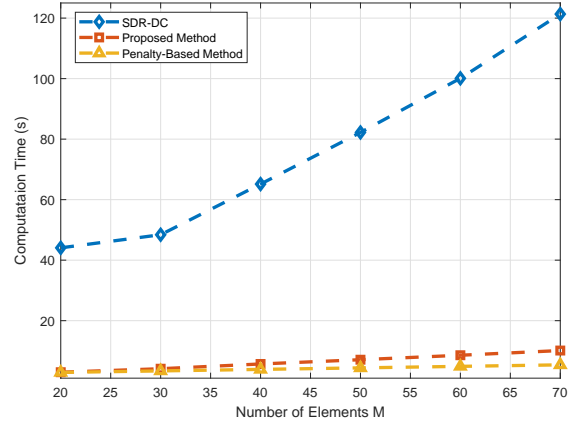}
        \caption{Average computation time versus $M$ with $K = 8$, $N = 10$.}
        \label{CT}
     \end{subfigure}
     \hfil
     \begin{subfigure}[b]{0.49\textwidth}
     \centering
  \includegraphics[scale=0.5]{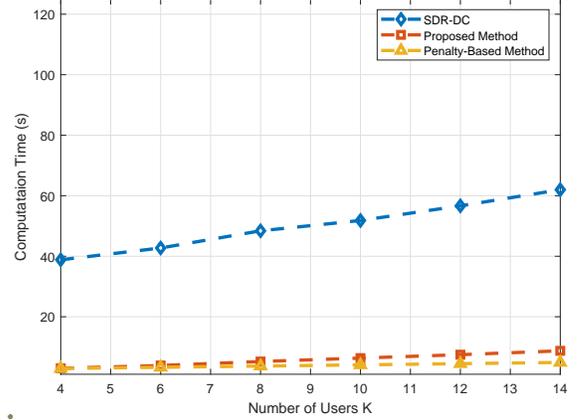}
        \caption{Average computation time versus $K$ with $M = 30$, $N = 10$.}
        \label{CTU}
     \end{subfigure}
        \caption{Average computation time compared with the SDR-DC and penalty-based method.}
        \label{fig:three graphs}
\end{figure}

\subsection{Performance Comparison with Other Schemes}

To verify the effectiveness of the STAR-RIS in the proposed MEC system and the performance of the proposed algorithm, we compare the performance with the following schemes.

\subsubsection{Conventional
RISs} In this case, the full-space coverage facilitated by the STAR-RIS in Fig. 1 is achieved by employing one conventional reflect-only RIS and one transmit-only RIS. The two conventional RISs are deployed adjacent to each other at the same location as the STAR-RIS. For a fair comparison, each conventional reflect-only/transmit-only RIS is assumed to have $M/2$ elements. This baseline scheme can be regarded as a special case of the STAR-RIS in MS mode, where $M/2$ elements can transmit the signal and $M/2$ elements can reflect the signal. 
\subsubsection{Random phase shift}
In this case, $\theta_m$ in $\boldsymbol{\Theta}$ are uniformly and randomly distributed in
$[0, 2\pi)$.
\subsubsection{Performance upper bound provided by energy splitting (ES)}

In this case, we assume that
all elements of the STAR-RIS are operated in the transmission and reflection mode, where the incident energy on each element is split into the energies of the transmitted and reflected signals with an energy splitting ratio of $ \rho^{\mathrm{t}}_M: \rho^{\mathrm{r}}_M$ and $\rho^{\mathrm{t}}_M, \rho^{\mathrm{r}}_M \in [0,1]$. The resulting optimization problem can be obtained by replacing the binary constraints (\ref{bi-sub}) with the constraints $\rho^{\mathrm{t}}_M, \rho^{\mathrm{r}}_M \in [0,1], \forall m \in \mathcal{M}$ in the problem (\ref{original-rho}) and can be solved by applying Algorithm 3. Notice that this scheme will theoretically obtain a better computation rate than the MS mode, as the $\boldsymbol{\rho}$ is relaxed from $\{0,1\}$ to $[0,1]$. However, as the ES mode cannot be implemented in hardware at this moment, this serves as an upper bound for the STAR-RIS.

\subsubsection{Equal time allocation}

In this case, we divide the length of the time slot $L$ equally into two parts and let the STAR-RIS transmits the signal half the time and reflects the signal half the time. This baseline scheme can be regarded as mode switching in the time domain.

\subsubsection{Zero-forcing (ZF) and Equal energy allocation}
The equal energy allocation scheme equally allocates the users' energy budget for local computing and computation offloading, and the ZF scheme leverages ZF receive beamforming at the AP. Note that the ZF receive beamforming cannot effectively deal with the cases when $N<K$, while our proposed optimal solution in Section IV can perform well even in these cases.

In Fig. \ref{Elements}, we show the computation rate of different schemes with respect to the number of elements of the STAR-RIS. From this figure, we can observe that all schemes' computation rates increase with the number of elements, which coincides with the intuition that STAR-RISs with more elements have a stronger capability to rectify the channel. It is clear that significant performance improvement can be achieved by the proposed BCD-optimized solution, verifying the great benefits of deploying STAR-RIS with joint optimization of the STAR-RIS's amplitude coefficients and phase shifts. It is confirmed that the STAR-RIS with MS mode provides a $17\%$ improvement in computation rate over the benchmark of conventional RISs and almost double in computation rate over the benchmark of random phase shift. In addition, the performance upper bound provided by ES mode outperforms MS mode by nearly $10\%$. We observed that the equal time allocation scheme only performs better than the random phase shift. This is because both users in the transmission or reflection space are served by STAR-RIS only for half the duration of the time length $L$.

\begin{figure}[t]
     \centering
    \includegraphics[scale=0.5]{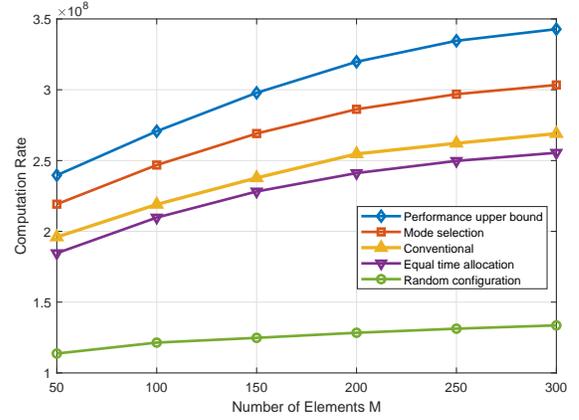}
        \caption{Computation rate versus the number of elements with $K = 8$, $N = 10$.}
        \label{Elements}
\end{figure}

The performance in terms of the computation rate versus the number of the AP’s antennas is presented in Fig. \ref{Antenna}. The system equipped with STAR-RIS always has better performance than the conventional RIS. Furthermore, as the number of antennas decreases, the performance of the ZF and equal energy allocation schemes degrade dramatically. It is because the ZF scheme cannot separate the signal streams when the number of users exceeds the number of antennas at the AP, while optimal beamforming design selects a beamformer between ZF and maximum ratio combining to maximize the SINR. Also, since the equal energy allocation scheme cannot control the uplink transmission power, if the number of antennas is less than the number of users (when the linear beamformer cannot eliminate other users' interference), it causes severe interference issues, and thus affects the offloading computation rates. In addition, we can observe that all the curves of the computation rate increase as $N$ grows, and the performance improvement becomes less prominent with larger $N$.

In Fig. \ref{Users}, we study the effects of the number of users, i.e., $K$, on the system performance of computation rate. It can be observed that the system equipped with STAR-RIS always performs better than the conventional RIS. Although the computation rates increase as the number of users grows, the performance improvement becomes less significant. It is due to the fact that the system is capacity limited when the number of users is small and becomes interference limited as the number of users increases. In addition, similar results can be observed from Fig. \ref{Antenna} that instead of performance degradation like ZF and equal energy allocation, the proposed solutions have better performance as $K$ becomes larger than $N$ through effectively designing the receive beamforming vectors and the users' energy allocation.

\begin{figure}[t]
     \centering
          \begin{subfigure}[b]{0.49\textwidth}
          \centering
  \includegraphics[scale=0.5]{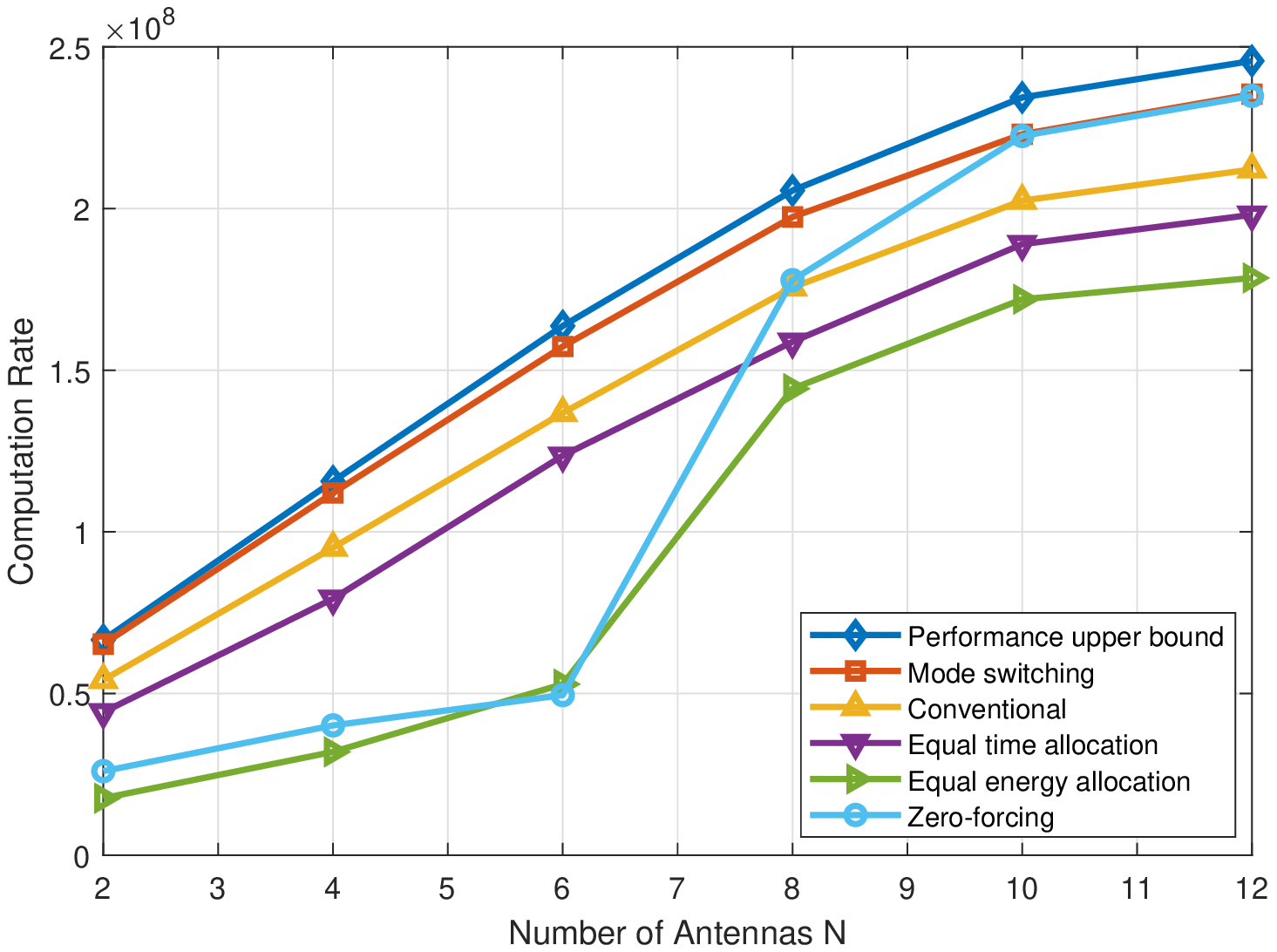}
\caption{}
        \label{Antenna}
     \end{subfigure}
          \hfil
     \begin{subfigure}[b]{0.49\textwidth}
         \centering
  \includegraphics[scale=0.5]{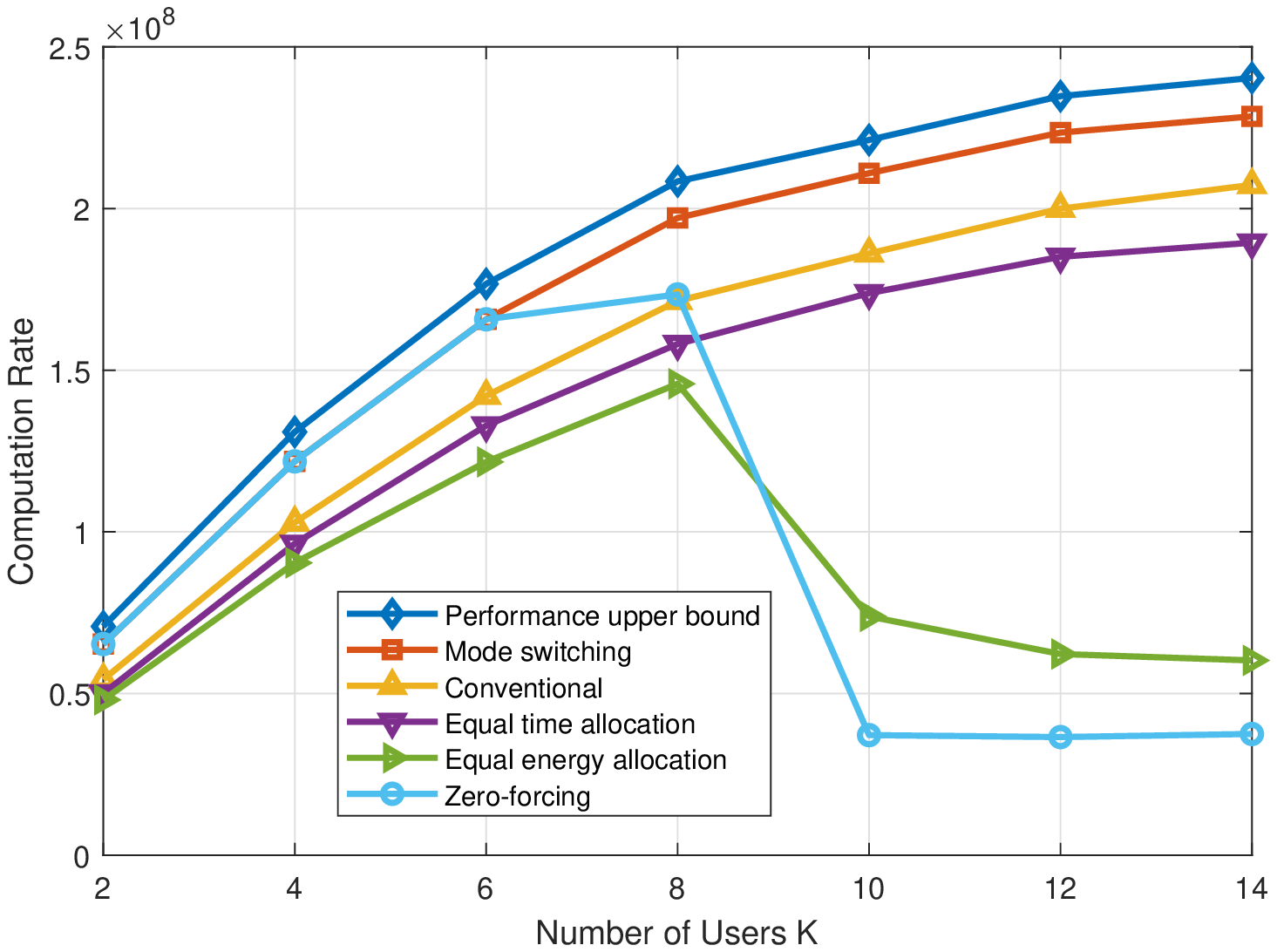}
        \caption{}
        \label{Users}
     \end{subfigure}
        \caption{Computation rate versus the number of users and number of antennas at the AP with $M=30$.}
\end{figure}

\begin{figure}[t]
     \centering
  \includegraphics[scale=0.5]{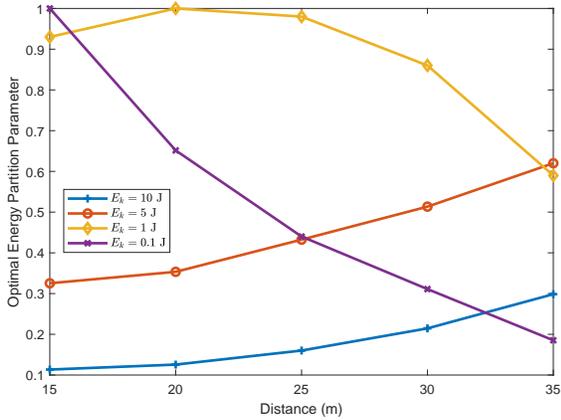}
        \caption{Optimal energy partition parameter versus distance with $M = 30$, $N = 10$.}
        \label{distance}
\end{figure}

To reveal insight into the energy trade-off between offloading and local computation, we investigate how the optimal energy partition parameter varies with the distance between STAR-RIS and the user. Specifically, in Figure \ref{distance}, we consider the user with varying energy budgets ranging from $E_k=0.1~\mathrm{J}$ to $E_k=10~\mathrm{J}$.
For users with high energy budgets, i.e., $E_k=5$ or $10~\mathrm{J}$, we observed that the optimal strategy is to allocate more energy to computation offloading as the distance increases. This is because computation offloading can provide a higher computation rate than local computation, and as the distance increases, the user needs to spend more energy on the data offloading.
In contrast, for a user with a low energy budget, i.e., $E_k=1~\mathrm{J}$, the optimal energy partition parameter first increases as the distance increases. However, as it reaches the maximum value of $1$, it starts to decline as the distance increases. The reason for the rise of the optimal energy partition parameter at the first segment is the same as that of users with a high energy budget. However, as the distance increases further, a limited energy budget cannot support high-speed data transmission in computation offloading; therefore, computation offloading becomes less economical than local computing. Thus, we observe a decline in the optimal energy partition parameter at the second segment. As for the case of $E_k=0.1~\mathrm{J}$, the optimal energy partition parameter reaches the maximum value of $1$ with a lower distance compared with $E_k=1~\mathrm{J}$ and then followed by a decline as the distance increases. The simulation results demonstrate that the optimal energy allocation strategy for computation offloading depends on the user's energy budget and the distance between the STAR-RIS and the user. In general, the optimal energy partition parameter first increases as the distance increases before reaching its maximum value of $1$ and then decreases as the distance increases. The optimal energy partition parameter curve will shift to the right if the energy budget is abundant, while the curve will shift to the left if the energy budget is small.

\section{Conclusion}
\label{con}
In this paper, a STAR-RIS-aided MEC system with computation offloading has been investigated. Specifically, the computation rate was maximized via the collaborative design of the STAR-RIS phase shifts, reflection and transmission amplitude coefficients, the AP's receive beamforming vectors, and the users' energy partition strategies for local computing and offloading. To handle the binary reflection and transmission amplitude coefficients of the STAR-RIS, the logarithm smoothing-based method has been proposed to eliminate the undesired stationary points and local optima induced by conventional penalty-based methods. Furthermore, a low-complexity iterative algorithm has been proposed to obtain a stationary point of the non-convex joint optimization problem. Due to the linear scaling of the computational complexity with respect to the number of users or STAR-RIS elements, the proposed algorithm is suitable for massive scenarios. Numerical results showed that the proposed low-complexity algorithm could save computation time by at least an order of magnitude compared to the DC programming and SDR-based solution. Besides, the STAR-RIS could significantly improve the computation rate of the system compared to the conventional RIS system. 


\section*{Appendix A\\
Degree of Freedom to Support $K$ Users with STAR-RIS}
We first examine the DoF between transmission users and the AP. The received signal  $\boldsymbol{y}_{\mathrm{t}} \in \mathbb{C}^{N\times 1}$ at the AP due to all transmission users is 
\begin{equation}
    \boldsymbol{y}_{\mathrm{t}} = \left(\boldsymbol{G}_{\mathrm{t}}\boldsymbol{\Theta}_{\mathrm{t}}\boldsymbol{H}_{\mathrm{s,t}}+\boldsymbol{H}_{\mathrm{d,t}}\right)\boldsymbol{s}_{\mathrm{t}}+\boldsymbol{z},
\end{equation}
where $\boldsymbol{H}_{\mathrm{d,t}} =[\boldsymbol{h}_{\mathrm{d},1} \  \cdots \ \boldsymbol{h}_{\mathrm{d},T}]\in\mathbb{C}^{N\times T}$ denotes the direct link between transmission users and the AP, $\boldsymbol{H}_{\mathrm{s,t}}\in\mathbb{C}^{Z \times T}$ denotes the link between transmission users and the STAR-RIS elements in the transmission mode, with $Z$ being the number of STAR-RIS elements in transmission mode, $\boldsymbol{\Theta}_{\mathrm{t}} \in \mathbb{C}^{Z \times Z}$ is the phase shift matrix of the STAR-RIS elements in transmission mode, $\boldsymbol{G}_{\mathrm{t}} \in \mathbb{C}^{N \times Z}$ is the channel from the elements in transmission mode to the AP,  and $\boldsymbol{s}_{\mathrm{t}} \in \mathbb{C}^{T\times 1}$ is the aggregated information symbols of transmission users. We here assume that the direct link $\boldsymbol{H}_{\mathrm{d,t}}$ has a rank $b>0$. The DoF of the equivalent channel $\boldsymbol{G}_{\mathrm{t}}\boldsymbol{\Theta}_{\mathrm{t}}\boldsymbol{H}_{\mathrm{s,t}}+\boldsymbol{H}_{\mathrm{d,t}}$ is its rank and is upper bounded by $\min \left(T, N, Z+b\right)$\cite{https://doi.org/10.48550/arxiv.2112.13787}. The intuition behind this result is that the direct path channel can be regarded as a separate RIS with $b$ elements where the phase shifts are fixed at 0 (or any arbitrary phase).

Similarly, the channel between reflection users to the AP can be written as $\boldsymbol{G}_{\mathrm{r}}\boldsymbol{\Theta}_{\mathrm{r}}\boldsymbol{H}_{\mathrm{s,r}}+\boldsymbol{H}_{\mathrm{d,r}}$, where $\boldsymbol{H}_{\mathrm{d,r}}=[\boldsymbol{h}_{\mathrm{d},T+1} \ \cdots \ \boldsymbol{h}_{\mathrm{d},K}]\in\mathbb{C}^{N\times R}$ with rank $j> 0$ denotes the direct link between reflection users and the AP, $R$ is the number of reflection users, $\boldsymbol{H}_{\mathrm{s,r}}\in\mathbb{C}^{F \times R}$ denotes the link between reflection users and the STAR-RIS elements in the reflection mode, with $F$ being the number of STAR-RIS elements in reflection mode, $\boldsymbol{\Theta}_{\mathrm{r}} \in \mathbb{C}^{F \times F}$ is the phase shift matrix of the STAR-RIS elements in reflection mode, and $\boldsymbol{G}_{\mathrm{r}} \in \mathbb{C}^{N \times F}$ is the channel from the STAR-RIS elements in reflection mode to the AP. Correspondingly, the rank of this channel is upper bounded by $\min \left(R, N, F+j\right)$.

Then the received signal at the AP due to all users can be represented as
\begin{equation}
\boldsymbol{y}=\underbrace{[\boldsymbol{H}_1 \  \boldsymbol{H}_2]}_{\boldsymbol{H}}\boldsymbol{s}+\boldsymbol{z},
\end{equation}
where $\boldsymbol{s} \in \mathbb{C}^{K\times 1}$ is the aggregated information symbols for all users, $\boldsymbol{H}_1=\boldsymbol{G}_{\mathrm{t}}\boldsymbol{\Theta}_{\mathrm{t}}\boldsymbol{H}_{\mathrm{s,t}}+\boldsymbol{H}_{\mathrm{d,t}}$, and $\boldsymbol{H}_2=\boldsymbol{G}_{\mathrm{r}}\boldsymbol{\Theta}_{\mathrm{r}}\boldsymbol{H}_{\mathrm{s,r}}+\boldsymbol{H}_{\mathrm{d,r}}$. Note that the maximum DoF for channel $\boldsymbol{H} \in \mathbb{C}^{N \times K}$ is not exactly $\min \left(N, K\right)$ but it is related to the DoF of $\boldsymbol{H}_1\in \mathbb{C}^{N \times T}$ and $\boldsymbol{H}_2 \in \mathbb{C}^{N \times R}$. To illustrate the above idea, we assume $N\geq K$ in the following. Since the maximum DoF of $\boldsymbol{H}_1$ is $\min \left(T, N, Z+b\right)$, if $Z+b < T$, then the DoF of $\boldsymbol{H}$ cannot reach $K$, since the rank of $\boldsymbol{H}_1$ is lower than $T$. Similarly, since the maximum DoF of $\boldsymbol{H}_2$ is $\min (T, N, F+j)$, if $F+j<R$, then the DoF of $\boldsymbol{H}$ cannot reach $K$ either, since the rank of $\boldsymbol{H}_2$ is lower than $R$. Therefore the necessary condition for $\boldsymbol{H}$ to have $K$ DoF to support $K$ users is the number of elements in transmission mode $Z$ greater than  or equal to $T-b$ and the number of elements in reflection mode $F$ greater than or equal to $R-j$. Since $Z+F=M$, we obtain $M \geq K-j-b$.

\section*{Appendix B\\
Derivation of the Gradient of $\nabla_{\boldsymbol{\rho}} \mathcal{Q}(\boldsymbol{\rho})$}
Firstly, the gradient of (\ref{smoothed}) for the third term $-\gamma \sum_m\sum_{\mathrm{x}}\left(\rho_{m}^{\mathrm{x}}-\left(\rho_{m}^{\mathrm{x}}\right)^{2}\right)$ and the fourth term $-\mu\Phi(\boldsymbol{\rho})$ with respect to $\rho_{m}^{\mathrm{x}}$ can be obtained as $-\gamma (1-2\rho_{m}^{\mathrm{x}})$ and $\mu\left(\frac{1}{\rho_m^{\mathrm{x}}}\!-\!\frac{1}{\left(1-\rho_m^{\mathrm{x}}\right)}\right)$, respectively. Then for the first term and second term: $\sum_{t=1}^{T}R_{t}(\boldsymbol{\rho})$ and $\sum_{r=1}^{R}R_{r}(\boldsymbol{\rho})$, the gradient with respect to $\rho_{m}^{\mathrm{x}}$ can be obtained based on chain rules for complex-valued variables\cite{petersen2008matrix}. The overall results are given below in (\ref{gradient_amp}).
\begin{figure*}
\begin{subequations}
\label{gradient_amp}
\begin{align}
\begin{split}
\nabla\!_{\rho_{m}^{\mathrm{t}}}\!\!\mathcal{Q}&(\rho_{m}^{\mathrm{t}})\!=\!-\!\!\sum_r\!\!\left(\!\!\frac{1}{\ln2(1\!+\!\gamma_{r})}2\!\operatorname{Re}\!\!\left(\!\!\frac{p_{r}\!\left|\left(\boldsymbol{v}_{r}\right)\!^{\mathrm{H}}\!
\boldsymbol{g}_{r}\!\right|^{2}\!\sum_{l \neq r, l \in\mathcal{R}}\!\!\left(p_{l}\boldsymbol{G}^{*}\boldsymbol{v}_{r}^{*}\boldsymbol{v}_{r}^{\mathrm{T}}\boldsymbol{g}_{l}^{*}\boldsymbol{h}_{\mathrm{r}, l}^{\mathrm{T}}\right)_{m,m}\!\!e^{j\theta_m}}{\left(\sum_{l \neq r, l \in \mathcal{R}} p_{l}\left|\left(\boldsymbol{v}_{r}\right)^{\mathrm{H}} \boldsymbol{g}_{l}
\right|^{2}\!+\!\sigma^{2}\left\|\boldsymbol{v}_{r}\right\|^{2}\right)^2}\!\!\right)\!\!\!\right)\!\!+\!\!\mu\!\left(\!\frac{1}{\rho_m^{\mathrm{t}}}\!-\!\frac{1}{\left(1-\rho_m^{\mathrm{t}}\right)}\!\!\right)\!\!-\!\gamma (\!1\!-\!2\rho_{m}^{\mathrm{t}})\\
&+\!\sum_t\!\left(\frac{1}{\ln2(1\!+\!\gamma_{t})}2\!\operatorname{Re}\!\left(\!\frac{\left(p_{t}\boldsymbol{G}^{*}\boldsymbol{v}_{t}^{*}\boldsymbol{v}_{t}^{\mathrm{T}}\boldsymbol{g}_{t}^{*}\boldsymbol{h}_{\mathrm{r}, t}^{\mathrm{T}}\right)_{m,m}\!\!e^{j\theta_m}}{\sum_{l \neq t, l \in \mathcal{T}} \!p_{l}\left|\left(\boldsymbol{v}_{t}\right)^{\mathrm{H}} \!\boldsymbol{g}_{l}\right|^{2}\!\!+\!\sigma^{2}\!\left\|\boldsymbol{v}_{t}\right\|^{2}}\!-\! \frac{p_{t}\!\left|\left(\boldsymbol{v}_{t}\right)^{\mathrm{H}}\!
\boldsymbol{g}_{t}
\right|^{2}\!\sum_{l \neq t, l \in \mathcal{T}}\!\left(p_{l}\boldsymbol{G}^{*}\boldsymbol{v}_{t}^{*}\boldsymbol{v}_{t}^{\mathrm{T}}\boldsymbol{g}_{l}^{*}\boldsymbol{h}_{\mathrm{r}, l}^{\mathrm{T}}\right)_{m,m}\!\!e^{j\theta_m}}{\left(\sum_{l \neq t, l \in \mathcal{T}} p_{l}\left|\left(\boldsymbol{v}_{t}\right)^{\mathrm{H}} \boldsymbol{g}_{l}
\right|^{2}\!+\!\sigma^{2}\left\|\boldsymbol{v}_{t}\right\|^{2}\right)^2}\right)\!\!\right),
\end{split}
\\
\begin{split}
\nabla\!_{\rho_{m}^{\mathrm{r}}}\!&\!\mathcal{Q}(\rho_{m}^{\mathrm{r}})\!=\!-\!\!\sum_t\!\!\left(\!\!\frac{1}{\ln2(1\!+\!\gamma_{t})}\!2\!\operatorname{Re}\!\!\left(\!\!\frac{p_{t}\!\left|\left(\boldsymbol{v}_{t}\right)\!^{\mathrm{H}}\!
\boldsymbol{g}_{t}\!
\right|^{2}\!\sum_{l \neq t, l \in\mathcal{T}}\!\left(p_{l}\boldsymbol{G}^{*}\boldsymbol{v}_{t}^{*}\boldsymbol{v}_{t}^{\mathrm{T}}\boldsymbol{g}_{l}^{*}\boldsymbol{h}_{\mathrm{r}, l}^{\mathrm{T}}\right)_{m,m}\!\!\!e^{j\theta_m}}{\left(\sum_{l \neq t, l \in \mathcal{T}} p_{l}\left|\left(\boldsymbol{v}_{t}\right)^{\mathrm{H}} \boldsymbol{g}_{l}
\right|^{2}\!+\!\sigma^{2}\left\|\boldsymbol{v}_{t}\right\|^{2}\right)^2}\!\!\right)\!\!\!\right)\!\!+\!\!\mu\!\left(\!\frac{1}{\rho_m^{\mathrm{r}}}\!-\!\frac{1}{\left(1-\rho_m^{\mathrm{r}}\right)\!}\!\right)\!\!-\!\gamma (1\!-\!2\rho_{m}^{\mathrm{r}})\\
&+\!\sum_r\!\left(\frac{1}{\ln2(1\!+\!\gamma_{r})}2\!\operatorname{Re}\!\left(\!\frac{\left(p_{r}\boldsymbol{G}^{*}\boldsymbol{v}_{r}^{*}\boldsymbol{v}_{r}^{\mathrm{T}}\boldsymbol{g}_{r}^{*}\boldsymbol{h}_{\mathrm{r}, r}^{\mathrm{T}}\right)_{m,m}\!\!e^{j\theta_m}}{\sum_{l \neq r, l \in \mathcal{R}} \!p_{l}\!\left|\left(\boldsymbol{v}_{r}\right)^{\mathrm{H}}\!\boldsymbol{g}_{l}
\right|^{2}\!\!+\!\sigma^{2}\!\left\|\boldsymbol{v}_{r}\right\|^{2}}\!-\! \frac{p_{r}\!\left|\left(\boldsymbol{v}_{r}\right)^{\mathrm{H}}\!
\boldsymbol{g}_{r}
\right|^{2}\!\sum_{l \neq r, l \in \mathcal{R}}\!\!\left(p_{l}\boldsymbol{G}^{*}\boldsymbol{v}_{r}^{*}\boldsymbol{v}_{r}^{\mathrm{T}}\boldsymbol{g}_{l}^{*}\boldsymbol{h}_{\mathrm{r}, l}^{\mathrm{T}}\right)_{m,m}\!\!\!e^{j\theta_m}}{\left(\sum_{l \neq r, l \in \mathcal{R}} p_{l}\left|\left(\boldsymbol{v}_{r}\right)^{\mathrm{H}} \boldsymbol{g}_{l}
\right|^{2}\!+\!\sigma^{2}\left\|\boldsymbol{v}_{r}\right\|^{2}\right)^2}\!\!\right)\!\!\right).
\end{split}
\end{align}
\end{subequations}
    \hrulefill 
\end{figure*}

\newpage

\section*{Appendix C\\
Proof of Proposition 1}
\label{appb}
Recall that the orthogonal projection of $\boldsymbol{\rho}$ is the optimal solution of (\ref{proj}), the Lagrangian of the problem is
\begin{equation}
\label{appbconvex}
L(\boldsymbol{\rho};\boldsymbol{\lambda})=\left\|\boldsymbol{\rho}-\boldsymbol{\rho}\left(i+\frac{1}{2}\right)\right\|^{2}+\sum_{m=1}^{M}\lambda_m\left(\boldsymbol{e}_m^{\mathrm{T}}\boldsymbol{\rho}-1\right),
\end{equation}
where $\boldsymbol{\lambda}$ refers to a collection of variables $\left\{\lambda_1, \ldots, \lambda_M\right\}$ and $\boldsymbol{e}_m \in \mathbb{R}^{2M} $ is a vector with its $m^{th}$ and $2m^{th}$ positions equal to 1 and others equal to 0. Since strong duality holds for the problem (\ref{appbconvex}), it follows that $\boldsymbol{\rho}^{\star}$ is an optimal solution to the problem (\ref{proj}) if and only if there exists $\boldsymbol{\lambda}^{\star} \in \mathbb{R}^{M}$ for which
\begin{subequations}
\begin{align}
\label{appb1}
\boldsymbol{\rho}^{\star} & \in \operatorname{argmin}_{\boldsymbol{0} \leq \boldsymbol{\rho} \leq \boldsymbol{1}} L(\boldsymbol{\rho};\boldsymbol{\lambda}^{\star}), \\
\label{appb2}
& \left[\begin{array}{cc}
\!\!I_{M},\!\!\!\!&I_{M}\!\!\!\! \\
\end{array}\right] \boldsymbol{\rho}^{\star} = \boldsymbol{1},
\end{align}
\end{subequations}
Using the expression of the Lagrangian given in (\ref{appbconvex}), the relation (\ref{appb1}) can be equivalently written as
\begin{equation}
    P_{\operatorname{Box}[\boldsymbol{0}, \boldsymbol{1}]}\left(\boldsymbol{\rho}\left(i+\frac{1}{2}\right)-\begin{bmatrix}
\boldsymbol{\lambda}^{\star} \\
\boldsymbol{\lambda}^{\star}
\end{bmatrix}\right).
\end{equation}
The feasibility condition (\ref{appb2}) can then be rewritten as
\begin{equation}
\left[\begin{array}{cc}
\!\!I_{M},&\!\!\!\!I_{M}\!\!\! \\
\end{array}\right]P_{\operatorname{Box}[\boldsymbol{0}, \boldsymbol{1}]}\biggl(\boldsymbol{\rho}\left(i+\frac{1}{2}\right)-\begin{bmatrix}
\boldsymbol{\lambda}^{\star} \\
\boldsymbol{\lambda}^{\star}
\end{bmatrix}\biggl)=\boldsymbol{1}.
\end{equation}

\section*{Appendix D\\
Proof of Proposition 2}
\label{appa}
First, by introducing a new variable $\eta_k$ to replace each ratio term inside the logarithm, (\ref{energyp}) can be rewritten as
\begin{subequations}
\label{13}
\begin{align}
\max _{\substack{\boldsymbol{a},\boldsymbol{\eta}}} \quad
&\sum_{k \in \mathcal{K}} \log_2 (\left.1+\eta_k\right)+\sum_{k \in \mathcal{K}}\frac{1}{C_{k}} \sqrt{\frac{(1-a_{k})E_{k}}{L\kappa_k}},\\
\text { s.t. } \quad & a_{k} \in[0,1], \quad \forall k \in \mathcal{K},\\
&\eta_k \leq \frac{p_{k}\left|\left(\boldsymbol{v}_{k}\right)^{\mathrm{H}}\!
\boldsymbol{g}_{k}
\right|^{2}}{\sum_{l \neq k} p_{l}\left|\left(\boldsymbol{v}_{k}\right)^{\mathrm{H}} \boldsymbol{g}_{l}
\right|^{2}\!+\!\sigma^{2}\left\|\boldsymbol{v}_{k}\right\|^{2}}, \quad \forall k \in \mathcal{K},
\end{align}
\end{subequations}
where $\boldsymbol{\eta}=\left\{\eta_1, \ldots, \eta_K\right\}$. The above optimization can be thought of as an outer optimization over $\boldsymbol{a}$ and an inner optimization over $\eta_k$ with fixed $\boldsymbol{a}$. The inner optimization is
\begin{subequations}
\label{convex}
\begin{align}
\max _{\substack{\boldsymbol{\eta}}} \quad
&\sum_{k \in \mathcal{K}} \log_2 (\left.1+\eta_k\right),\\
\label{SINR-con}
\text { s.t. } \quad 
&\eta_k \leq \frac{p_{k}\left|\left(\boldsymbol{v}_{k}\right)^{\mathrm{H}}\!
\boldsymbol{g}_{k}
\right|^{2}}{\sum_{l \neq k} p_{l}\left|\left(\boldsymbol{v}_{k}\right)^{\mathrm{H}} \boldsymbol{g}_{l}
\right|^{2}\!+\!\sigma^{2}\left\|\boldsymbol{v}_{k}\right\|^{2}}, \quad \forall k \in \mathcal{K}.
\end{align}
\end{subequations}
Note that (\ref{convex}) is a convex optimization problem, so the strong duality holds. We introduce the dual variable $\zeta_k$ for each inequality constraint in (\ref{SINR-con}) and form the Lagrangian function
\begin{equation}
\label{lagrang}
\begin{aligned}
\mathcal{L}&(\boldsymbol{\eta},\boldsymbol{\zeta})\!=\!\sum_{k \in \mathcal{K}} \log_2 (\left.1+\eta_k\right)\\
&\!-\sum_{k \in \mathcal{K}}\!\zeta_k\!\left(\!\eta_k\!-\!\frac{p_{k}\left|\left(\boldsymbol{v}_{k}\right)^{\mathrm{H}}\!
\boldsymbol{g}_{k}
\right|^{2}}{\sum_{l \neq k} \!p_{l}\!\left|\left(\boldsymbol{v}_{k}\right)^{\mathrm{H}}\!\boldsymbol{g}_{l}
\right|^{2}\!\!\!+\!\sigma^{2}\!\left\|\boldsymbol{v}_{k}\right\|^{2}}\right),
\end{aligned}
\end{equation}
where $\boldsymbol{\zeta}$ refers to the collection of variables $\left\{\zeta_1, \ldots, \zeta_K\right\}$. Due to strong duality, the optimization (\ref{convex}) is equivalent to the dual problem
\begin{equation}
\label{15}
\min_{{\boldsymbol{\zeta}} \succeq 0} \ \max_{\boldsymbol{\eta}} L(\boldsymbol{\eta}, \boldsymbol{\zeta}).
\end{equation}
Let $\left(\boldsymbol{\eta}^{\star}, \boldsymbol{\zeta}^{\star}\right)$ be the saddle point of the problem (\ref{15}). It must satisfy the first-order condition $\partial L / \partial \eta_k=0$ :
\begin{equation}
    \zeta_k^{\star}=\frac{1}{(1+\eta_k^{\star})\ln 2}, \quad \forall k \in \mathcal{K}.
\end{equation}

Given the constraint (\ref{SINR-con}), the optimal $\eta_k^{\star}$ is obtained with the equality of (\ref{SINR-con}). Hence, the optimal $\zeta_k^{\star}$ is given by
\begin{equation}
\label{17}
\zeta_k^{\star}=\frac{1}{\ln 2}\frac{\sum_{l \neq k} p_{l}\left|\left(\boldsymbol{v}_{k}\right)^{\mathrm{H}} \boldsymbol{g}_{l}
\right|^{2}\!+\!\sigma^{2}\left\|\boldsymbol{v}_{k}\right\|^{2}}{\sum_{l \in \mathcal{K}} p_{l}\left|\left(\boldsymbol{v}_{k}\right)^{\mathrm{H}} \boldsymbol{g}_{l}
\right|^{2}\!+\!\sigma^{2}\left\|\boldsymbol{v}_{k}\right\|^{2}}, \quad \forall k \in \mathcal{K}.
\end{equation}
Note that $\zeta_k^{\star} \geq 0$ is automatically satisfied since both the numerator and denominator in (\ref{17}) are positive. Putting (\ref{17}) in (\ref{15}), problem (\ref{13}) can then be reformulated as
\begin{equation}
\label{final}
\max_{\boldsymbol{\eta}} L\left(\boldsymbol{\eta}, \boldsymbol{\zeta}^{\star}\right).
\end{equation}
Furthermore, combining with the outer maximization over $\boldsymbol{a}$ and after some algebra, we can find (\ref{final}) to be the same as the maximization of (\ref{DD}) in the Proposition \ref{prop1}.

\bibliographystyle{IEEEtran}
\bibliography{IEEEabrv,my}

\end{document}